\documentclass[%
 reprint,
superscriptaddress,
 amsmath,amssymb,
 aps,
]{revtex4-2}
\usepackage{graphicx}
\usepackage{dcolumn}
\usepackage{bm}
\usepackage[caption=false]{subfig}
\usepackage{amsmath}
\usepackage{comment}
\usepackage{calligra}
\DeclareMathAlphabet{\mathcalligra}{T1}{calligra}{m}{n}
\DeclareFontShape{T1}{calligra}{m}{n}{<->s*[2.2]callig15}{}

\usepackage{amssymb}
\usepackage{tabularx}
\usepackage{multirow}  
\usepackage{xcolor}
\usepackage{slashed}
\usepackage{url}
\usepackage{xcolor}
\definecolor{mycolor}{RGB}{0,0,204}
\usepackage{hyperref}
\hypersetup{
  colorlinks   = true,    
  urlcolor     = blue,    
  linkcolor    =blue,    
  citecolor    = blue      
}
\usepackage{graphicx}
\usepackage{dcolumn}
\usepackage{bm}

\begin{document}

\preprint{APS/123-QED}

\title{New Inflation in Waterfall Region} 
\author{Niamat Ullah Khan}%
\email{niamatullahk7@gmail.com}%
\affiliation{Department of Physics, Quaid-i-Azam University, Islamabad 45320, Pakistan
}
\affiliation{Department of Physics, Balochistan University of Information Technology, Engineering and Management Sciences, Quetta 87300, Pakistan}
\author{Nadir Ijaz}%
 \email{nadirijaz1919@gmail.com}%
\affiliation{Department of Physics, Quaid-i-Azam University, Islamabad 45320, Pakistan 
}%
\author{Mansoor Ur Rehman}%
 \email{mansoor@qau.edu.pk}
\affiliation{Department of Physics, Quaid-i-Azam University, Islamabad 45320, Pakistan 
}%



\begin{abstract}
We introduce a class of new inflation models within the waterfall region of a generalized hybrid inflation framework. The initial conditions are generated in the valley of hybrid preinflation. Both single-field and multi-field inflationary scenarios have been identified within this context. A supersymmetric realization of this scenario can successfully be achieved within the tribrid inflation framework.
To assess the model's viability, we calculate the predictions of inflationary observables using the $\delta N$ formalism, demonstrating excellent agreement with the most recent Planck data. Furthermore, this model facilitates successful reheating and nonthermal leptogenesis, with the matter-field component of the inflaton identified as a sneutrino.
\end{abstract}

\maketitle


\section{\label{sec:level1}Introduction}
The paradigm of accelerated expansion in the early universe \cite{PhysRevD.23.347} has been a very interesting idea because it elegantly resolves the horizon and the flatness problems in the standard big-bang model of cosmology \cite{Linde:1981mu, PhysRevLett.48.1220}. Numerous models have been proposed to describe this inflationary phase in the early universe \cite{linde2005particle,Lyth:1998xn,Martin:2013tda,Mazumdar:2010sa}. One prominent model in this context is the `new inflation' model \cite{linde2005particle,Linde:1981mu,PhysRevLett.48.1220}, which belongs to the category of small-field inflation models and aligns well with the effective field theory description. This model typically emerges as a result of spontaneous symmetry breaking, usually associated with a grand unified theory \cite{Linde:1981mu, Albrecht:1982wi,Rehman_2020} or a flavor symmetry \cite{Antusch_2008}. Moreover, the prospect of achieving a low reheat temperature is an appealing aspect, particularly in the context of a supergravity realization of new inflation \cite{Izawa_1997}, as it helps avoid the gravitino overproduction problem~\cite{ELLIS1984181,Khlopov:1984pf}.

Nonetheless, the new inflation model faces a serious challenge. In order for new inflation to last long enough, the initial condition must be extremely fine-tuned. Essentially, the inflaton field $\phi$ must be smooth over a substantial region with an average value that is very close to a local maximum of the potential $V(\phi)$. There should be some dynamical mechanism guiding the universe to select such a specific value for the inflaton field $\phi$. In simpler terms, this is a quest to find the answer to the question: ``Who put it there?''~\cite{baumann_2022}.

In order to dynamically elucidate the initial conditions for `new inflation', a prior inflationary phase known as 'pre-inflation' can be introduced before the onset of 'new inflation.'
This idea of pre-inflation was successfully realized within the framework of supergravity in \cite{Izawa_1997,Kawasaki:1998vx,Kanazawa:1999ag,Yamaguchi:2004tn,_eno_uz_2004}, where it typically entails the incorporation of a new set of gauge singlet fields.
For a specific realization of pre-inflation within the context of the supersymmetric tribrid inflation models, where the inflaton is primarily coupled to matter fields, please refer to \cite{Antusch_2014}.

A new inflationary phase has been recently identified within the waterfall region of a hybrid inflation model \cite{Clesse_2011,Kodama_2011}, where the standard hybrid inflation \cite{Linde_1994} is employed as a pre-inflationary phase. The initial conditions for this new inflationary phase, commonly known as `waterfall inflation', are dynamically generated through quantum fluctuations occurring in the waterfall field around the critical point of the waterfall transition. It is worth noting that while this waterfall inflation model does yield a scalar spectral index that is red-tilted, it does not align with the latest Planck data ~\cite{Plank_2018}.

In the present article, we have developed a model of new inflation as a generalization of the original waterfall inflation model \cite{Clesse_2011}. This model is designed to yield predictions for inflationary observables that are consistent with the latest Planck data.
The initial conditions for this waterfall inflation are generated via a diffusion boundary, which arises from quantum fluctuations in the waterfall field around the critical point of the waterfall transition. 
In the non-supersymmetric version of our model, the scalar potential's form can be constrained by a global symmetry denoted as $Z_{p}\times Z_{q}$. 
Furthermore, for a potential supersymmetric realization, we employ the waterfall region within the tribrid inflation model. A scalar matter field, such as the sneutrino, can play the role of the inflaton here. We calculate the inflationary predictions for a generic GUT model, incorporating a consistent framework for reheating and leptogenesis. 

The paper is structured as follows: In section \ref{sec:level2}, 
we provide a brief overview of the original hybrid inflation model, focusing on waterfall inflation. We also derive analytical expressions for inflationary observables within the new inflation limit, which will be essential for our subsequent discussions. In the next section \ref{sec:level3},
we introduce a generalized version of the model aimed at addressing the challenges encountered in the original model. Building on the insights from Section \ref{sec:level2}, we analyze our model by predicting inflationary observables consistent with the range of measured data. Section \ref{sec-4}
delves into the supersymmetric realization of the model. We provide a comprehensive analysis of this realization along with a coherent scenario for reheating and leptogenesis. Finally, in section \ref{conclusion}, we summarize our findings and conclusions.
 
\section{\label{sec:level2}Original Hybrid Inflation}
\begin{figure}[t]
\includegraphics[width=70mm,scale=0.7]{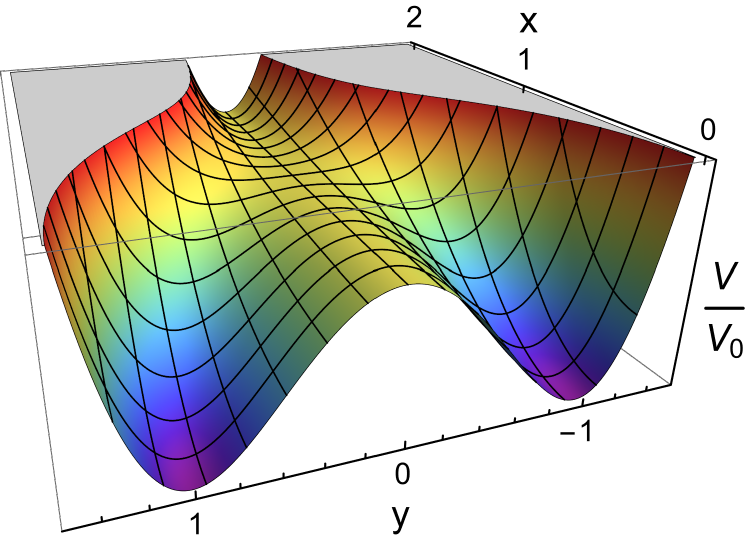}%
\caption{\label{Fig1}  The normalized scalar potential $V/V_0$  as a function of $x=\phi/\phi_c$ and $y=\psi/M$ for $\phi_{c}=M$. Here, $V_0=\Lambda^4$.}
\end{figure}

In this section, we provide a quick overview of the original hybrid inflation model as initially proposed in \cite{Linde_1994}, followed by a discussion of the subsequent advancements in the waterfall inflationary phase \cite{Clesse_2011}, which aligns with our suggested generalization.
The scalar potential for this model can be written as,
\begin{equation} \label{1}
\small V(\phi,\psi)=\Lambda^4 \left[\left(1-\left(\frac{\psi}{ M}\right)^2\right)^2 +\left(\frac{\phi}{\mu}\right)^2+2 \left(\frac{\phi}{\phi_{c}}\right)^2\left(\frac{\psi}{M}\right)^2\right],
	\end{equation}
where the scalar fields $\phi$ and $\psi$ represent the inflaton and waterfall-Higgs fields, while $M$ and $\mu$ denote the mass parameters. The potential parameter $\Lambda$ determines the energy scale of inflation. The above form of the potential can be constrained by a discrete symmetry $Z_2$ where both $\phi$ and $\psi$ are odd under this symmetry.
This scalar potential is displayed in Fig.~\ref{Fig1} where a ﬂat valley with $\psi = 0$ and $\phi > \phi_c$, suitable for inﬂation, is clearly visible. 

In the original version of the model, inflation takes place within the $\psi=0$ valley when $\phi > \phi_c $, where $\phi_{c}$ represents the critical value of $\phi$ below which the mass of $\psi$ becomes tachyonic. 
Within the framework of this effective single-field paradigm, inflation is assumed to terminate abruptly when the inflaton field falls below its critical value, $\phi_c$. Following this, both fields rapidly stabilize at their respective global minima, $\phi=0$ and $\psi=M$. This version is commonly referred to as the original hybrid inflation, also known as valley inflation \cite{Linde_1994}.

The end of the inflationary epoch and the subsequent breaking of the underlying symmetry can potentially give rise to topological defects, such as domain walls when the $Z_2$ symmetry is broken. The model's initial prediction for the spectral index, $n_s$, tends to exhibit a blue tilt. However, by including one-loop radiative corrections originating from inflaton couplings, which are essential for reheating, the value of $n_s$ can be adjusted to be consistent with observations \cite{Rehman_2009, Ahmed_2016}. For a supersymmetric realization of this standard version, please refer to the following references: \cite{PhysRevLett.73.1886,Copeland_1994,Buchm_ller_2000,Senoguz_2003,_eno_uz_2004,_eno_uz_2005,Jeannerot_2006,Bastero_Gil_2007,ur_Rehman_2007,Clesse_2009,Rehman_2010,REHMAN201075,Shafi_2011,Rehman_2011,Civiletti_2011,Buchm_ller_2014,Buchm_ller_2021,Ahmed_2022,Ahmed_2023,afzal2023supersymmetric}.	
For a similar waterfall structure in the tribrid inflation models, see ~\cite{Antusch_2005,Antusch_2010,Antusch_2012, Antusch:2012jc}. It is pointed out in \cite{Clesse_2011} that the waterfall dynamics of hybrid inflation might change for some cases and the inflation can still occur with a large number of efolds even during the waterfall regime ($\phi<\phi_c$ with $\psi \neq 0$). For a more detailed analytical study see Ref.~\cite{Kodama_2011}.

The waterfall phase can be further divided into two distinct regimes: the mild waterfall and the longer waterfall. 
The mild waterfall requires a multifield treatment and corresponds to a relatively small number of e-folds, typically when $N \gtrsim 60$. However, a recent finding ~\cite{Clesse_2014} suggests that it is challenging to attain the correct values for both the spectral index and the amplitude of the scalar power spectrum  simultaneously while remaining consistent with observational constraints.

The longer waterfall phase emerges as an effective single-field inflationary model, encompassing a substantial number of e-folds, $N \gg 60$. This characteristic aligns it with inflationary scenarios such as new inflation or type-I hilltop inflation \cite{Linde:1981mu, Boubekeur_2005,Kohri_2007,Antusch_2008}. In this regime, the topological defects resulting from symmetry breaking are rapidly inflated away.
It is important to note that in this new inflation limit, a red-tilted scalar spectral index \cite{Clesse_2014}, given by $n_s\simeq1-\frac{4}{N_0}$, is obtained. This is in contrast to the blue-tilted spectral index predicted by the original hybrid inflation model. However, for a realistic range of $N_0$ spanning from 50 to 60, its value remains around $\simeq 0.92 - 0.94$, which does not align with observational data.
This is briefly reviewed in this section.

\subsection*{Fields Dynamics and Initial Conditions} 
The background equations of motion for the scalar  fields are,
	\begin{equation}\label{2}
		\ddot{\phi} + 3H\dot{\phi} +\partial_{\phi}  V= 0,\quad
		\ddot\psi + 3H\dot\psi +\partial_\psi  V= 0,
	\end{equation}
with 
	\begin{equation}\label{3}
		H^2= \frac{1}{3}\left( \frac{\dot{\phi}^2}{2} + \frac{\dot{\psi}^2}{2} + V(\phi, \psi) \right), 
	\end{equation}
where $\partial_{\phi} V \equiv \partial V/ \partial {\phi}$, $\partial_\psi V \equiv \partial V/\partial \psi$. Throughout the text, both here and in subsequent sections, we will employ Planckian units, where the reduced Planck mass is taken as unity, $m_P=1$. Under the slow-roll approximation ($ \dot{\phi}^2, \dot{\psi}^2\ll V(\phi,\psi), \ddot{\phi}\ll 3H\dot{\phi},\partial_{\phi} V $ and  $\ddot{\psi}\ll 3H\dot{\psi},\partial_\psi V $), the equations of motion presented above can be simplified for the scalar potential defined in Eq.~(\ref{1}) as
\begin{align}\label{aaa}
\phi^{\prime}(N)&\simeq -\frac{2\phi}{\mu^{2}}\left(1+\frac{2\mu^2}{\phi^{2}_{c}}\left(\frac{\psi}{M}\right)^{2}\right), \\
  \psi^{\prime}(N)&\simeq \frac{4\psi}{M^{2}}\left[\left(1-\left(\frac{\phi}{\phi_{c}}\right)^{2}\right)-\left(\frac{\psi}{M}\right)^{2} \right],
  \label{256}
	\end{align}
where primes are derivatives with respect to the number of efolds with $\dot{\phi} = H\frac{d\phi}{dN}$, and $H^2\simeq\frac{\Lambda^4}{3}$.

To delve into the initial conditions of the `new inflationary phase', our first step is to examine the field dynamics preceding the waterfall transition, where $\phi > \phi_{c}$. In this particular regime, we make the assumption that the waterfall field remains settled at zero far from the critical point at $\phi_{c}$. The field dynamics in this region primarily remain classical until we approach the quantum diffusion region near the waterfall point, where quantum fluctuations in the waterfall field eventually dominate over its classical motion. 

To describe this quantum diffusion process, the Langevin equation \cite{Garc_a_Bellido_1996} can be employed to calculate the quantum dispersion, denoted as $\langle\psi^{2} \rangle$, in the waterfall field. The square root of this dispersion is taken as the initial value of $\psi$ around the critical point, $\phi \simeq \phi_c$, with $\langle \psi \rangle =0$ \cite{Clesse_2011},
\begin{equation}\label{ii6}
 \psi_{i} \equiv \sqrt{\langle\psi^{2} \rangle} \simeq \left( \frac{ \Lambda^{8} \mu \, M}{96 \, \pi^{\frac{3}{2}}} \right)^{\frac{1}{2}}.
\end{equation}
Consequently, the system gets off the valley and starts rolling down in the waterfall region. This generic quantum displacement in the waterfall field generates the initial conditions of the `new inflationary phase' in the waterfall region as discussed below.

\subsection*{New Inflation Limit}
As outlined in the refs.~\cite{Clesse_2011,Kodama_2011}, the evolution of the scalar fields in the waterfall region can be divided into three distinct phases, aptly named phase-0, phase-1, and phase-2, following a chronological sequence.
In phase-0, which marks the initial stage of the waterfall regime, the field $\phi$ is approximately equal to $\phi_{c}$, and $\psi$ is significantly smaller. To be more specific, we have $\frac{\sqrt{2}\mu\psi}{\phi_{c}M}\ll 1$. 
During this phase, the number of efolds, $N_{0}\simeq \frac{\mu^{2}}{4}\left(\frac{\psi_{i}}{M}\right)^{2}$, is very small and the phase ends quickly.

After the end of phase-0, $\phi$ undergoes a gradual decrease and begins to deviate from $\phi_{c}$.
This transition marks the onset of phase-1, where the first term in Eq.~(\ref{256}) becomes dominant.
While $\psi$ experiences an increase in this phase, the second term in Eq.~(\ref{aaa}) remains sufficiently small to be disregarded.  
A mild waterfall inflation with $N \gtrsim 60$ can be readily achieved in this phase by selecting a sufficiently large value for $\mu$, given a specific value for $M$ and $\phi_c$. 
By opting for an even greater value of $\mu$ but still keeping $\frac{\sqrt{2}\mu\psi}{\phi_{c}M} < 1$, it becomes possible to achieve long waterfall inflation with $N \gg 60$. 
Nonetheless, the scalar power spectrum adopts a Gaussian shape with its peak at $\phi_c$, making it challenging to meet both the amplitude and tilt constraints imposed by the latest Planck data \cite{Plank_2018} simultaneously.

In phase-2, the equations of motion (\ref{aaa}) and (\ref{256}) are primarily described by their respective second and first terms. The critical value $\psi_2 = \frac{\phi_c M}{\sqrt{2} \mu}$ serves as the boundary between phase-1 and phase-2. 
Given a GUT scale value of $\phi_c \sim M$, achieving phase-2 typically requires a large transPlanckian value for $\mu$,  
deep within phase 2, when $\frac{\sqrt{2}\mu\psi}{\phi_{c}M} \gg 1$, a phase of effective single-field inflation emerges and is recognized as `new inflation'. While this phase has been previously discussed in \cite{Kodama_2011, Clesse_2014, Clesse_2012}, our presentation here describes the 'new inflation limit' in a manner that aligns with our subsequent discussions and potential generalizations.

The equations of motion for scalar fields in phase-2 can be simplified as:
\begin{equation}\label{6}
	\phi^{\prime}(N)\simeq- \frac{4 \phi}{\phi^{2}_c} \left(\frac{\psi}{M}\right)^{2},\quad
 \psi^{\prime}(N)\simeq\frac{4}{M} \left(\frac{\psi}{M}\right)\left(1-\frac{\phi^2}{\phi^{2}_c}\right).
\end{equation}
Combining these two equations, we arrive at the following equation,
\begin{equation}\label{8}
	\frac{dX}{d\chi} \simeq -\left(\frac{X}{1-X^{2}}\right)\chi, 
\end{equation}
where, for convenience, we introduce the normalized fields, $X = \phi / \phi_{c}$ and $\chi = \psi / \phi_{c}$. In the small $\chi$ limit, the above equation admits the following approximate solution,
\begin{equation}\label{11}
	X \simeq  \left(1-\frac{1}{\sqrt{2}}\left(X - X_2 \right)\right),
\end{equation}
where $X_2 = \psi_2 / \phi_c$. This expression exhibits an effective single-field behavior, particularly in the regime of new inflation, where $\psi \gg \psi_2$, making the influence of $\psi_2$ negligible in subsequent calculations. Therefore, we consider the following expression for our later discussion,
\begin{equation}\label{11}
	\phi \simeq \phi_{c}\left(1-\frac{1}{\sqrt{2}}\left(\frac{\psi}{\phi_{c}}\right)\right).
\end{equation}

The joint dynamics of the fields $\phi$ and $\psi$ can be effectively described in terms of an adiabatic field, $\sigma$, defined by the equation:
\begin{equation}\label{Gxv2}
   \Dot{\sigma}  =\cos{\theta}\Dot{\phi}+\sin{\theta}\Dot{\psi},
 \end{equation}
where the coefficients $\cos{\theta}$ and $\sin{\theta}$ are determined as follows:
\begin{align}
    \cos{\theta}=\frac{\Dot{\phi}}{\sqrt{\Dot{\phi}^{2}+\Dot{\psi}^{2}}},\qquad
    \sin{\theta}=\frac{\Dot{\psi}}{\sqrt{\Dot{\phi}^{2}+\Dot{\psi}^{2}}}.
\end{align}
From Eq.~(\ref{11}), we can approximate $\frac{d\phi}{d\psi}\simeq -\frac{1}{\sqrt{2}}$, and we have $\cos{\theta}\simeq -\frac{1}{\sqrt{3}}$ and $\sin{\theta}\simeq \sqrt{\frac{2}{3}}$. Consequently, Eq.~(\ref{Gxv2}) implies that $\sigma \simeq \sqrt{\frac{3}{2}}\psi$. This identification allows us to express the potential in (\ref{Gi}) in terms of the adiabatic field $\sigma$ as follows:
\begin{equation}\label{Gxvi}
		V\left(\sigma\right) \simeq \Lambda^{4}\left[ 1-\frac{\mathcal{\alpha}}{\sqrt{2}} \sigma^{3}  \right],
\end{equation} 
where $\alpha\equiv \frac{4}{M^{2}\phi_{c}\left( \sqrt{\frac{3}{2}} \right)^3}$. 
With this parameterization, the potential in Eq.~(\ref{1}) assumes a standard new inflation model form. In the subsequent subsection, we will derive its predictions within the framework of the slow-roll approximation.

\subsection*{The Slow-roll Predictions}
The first and second slow-roll parameters are,
\begin{equation}\label{i}
\epsilon\simeq\frac{1}{2}\left( -\frac{3\mathcal{\alpha}}{\sqrt{2}}\sigma^{2}\right)^{2}, \qquad \eta \simeq  -3\sqrt{2}\mathcal{\alpha}\sigma .
\end{equation}
As $\epsilon$ is negligibly small, the inflation ends when $\eta(\sigma_{e}) \simeq -1$. Solving for the field value at the end of inflation, we find:
\begin{equation}\label{sige}
\sigma_{e} \simeq \frac{1}{3\sqrt{2}\mathcal{\alpha}}.
\end{equation}
The number of e-folds, $N_0$, before the end of inflation is given by,
\begin{equation}\label{iii}
N_{0} \simeq \int_{\sigma_{e}}^{\sigma_{0}}\left(\frac{V}{V_{\sigma}}\right)d{\sigma},
	\end{equation}
where $\sigma_0$ corresponds to the field value that normalizes the scalar power spectrum,
\begin{equation}\label{iv3}
A_s(k_0) = \frac{1}{24\pi^2} \frac{V}{\epsilon(\sigma_0)} \simeq \frac{1}{54\pi^2}\frac{\Lambda^4}{\alpha^2 \sigma_0^4},
\end{equation}
to Planck's measurement, $A_s(k_0) =2.4 \times 10^{-9} $, at the pivot scale $k_0 = 0.05$~Mpc$^{-1}$.
Using Planck's normalization, the value of $\Lambda$ can be calculated in terms of $\sigma_0$, which can, in turn, be expressed in terms of $N_0$ using (\ref{sige}), 
\begin{equation}\label{iv3}
\sigma_{0} \simeq \frac{\sqrt{2}}{3\alpha}\frac{1}{N_0+2}.
\end{equation}

The scalar spectral index, $n_s$, can now be written in terms of the e-folds $N_{0}$ using Eq.~(\ref{iv3}), 
\begin{equation}\label{v}
n_{s}  \simeq  1 + 2 \eta \simeq 1-\frac{4}{N_{0}},
\end{equation}
while the tensor-to-scalar ratio $r$ is given by,
\begin{equation}\label{vi}
 \begin{split}
r & = 16 \, \epsilon\simeq 
8\left(-\frac{\sqrt{2}}{3\alpha}\frac{1}{\left(N_0+2\right)^2}\right)^2.
\end{split}
\end{equation}
For a typical range of efolds, $N_0 \simeq 50-60$, we obtain $n_s \simeq 0.92-0.94$, $r \simeq (3-5) \times 10^{-20}$, $\Lambda \simeq 2 \times 10^{-7}$ and $\sigma_0 \simeq 4 \times  10^{-9}$ assuming $\phi_c = M = 0.01$.
However, it's worth noting that the scalar spectral index value obtained here appears to be too red-tilted to align with the Planck 2018 data \cite{Plank_2018}. In the subsequent sections, we will extend this new inflation model to accommodate the Planck constraint on the scalar spectral index.

\section{\label{sec:level3}General Case}
\begin{figure}[t!]\centering
\subfloat[\label{2a}]
{\includegraphics[width=0.3\textwidth]{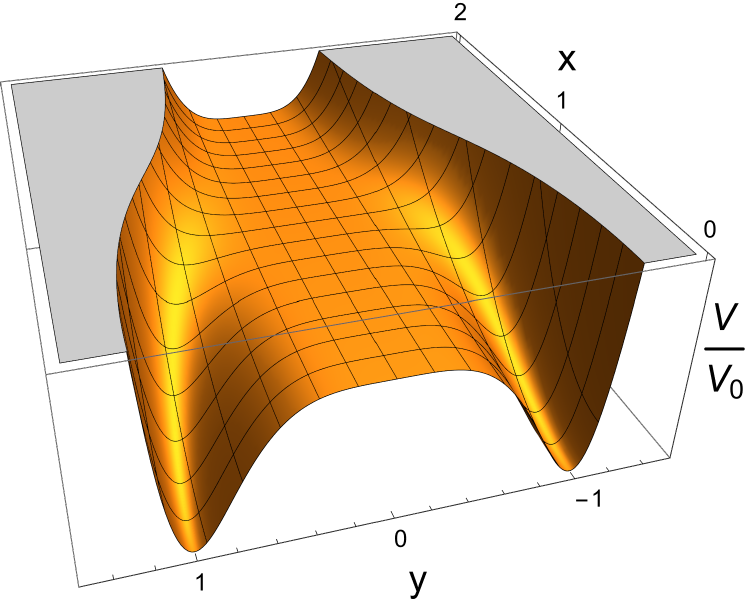}}\qquad
\subfloat[\label{2b}]
{\includegraphics[width=0.3\textwidth]{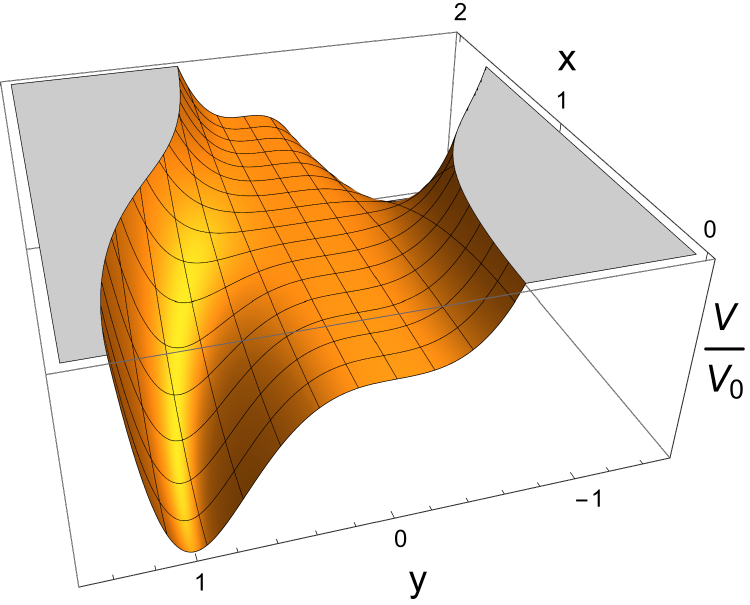}}
\caption{\label{Fig2}The normalized scalar potential $V/V_0$ as a function of $x=\phi/\phi_c$ and $y=\psi/M$, for (a)~$q=4$ and (b)~$q=3$.}

\end{figure}

For a possible generalization of the potential in Eq.~(\ref{1}) which can lead to single-field inflation in the large $\mu$ limit we consider the following form of the potential,
\begin{equation} \label{Gi}
\small V(\phi,\psi)=\Lambda^4 \left[\left(1-\left(\frac{\psi}{ M}\right)^q\right)^2 +\left(\frac{\phi}{\mu}\right)^p+2 \left(\frac{\phi}{\phi_{c}}\right)^p\left(\frac{\psi}{M}\right)^q\right],
\end{equation}
where $p$ and $q$ are integers. In order to have a stable valley before the waterfall region $q$ should be even (see, e.g. Fig.~\ref{2a} for $q=4$), otherwise, the concavity of the valley for a given value of $\phi$ changes around the $\psi =0$ point which acts as a point of inflection (See, e.g. Fig.~\ref{2b} for $q=3$). For increasing values of $q$ the flatness of the potential near the critical point increases which leads to a large number of efolds. This particular form of the potential can be constrained by a global symmetry such as $Z_{p}\times Z_{q}$.
Under this symmetry $\psi$ ($\phi$) carries a unit charge under $Z_q$ ($Z_p$) while $\phi$ ($\psi$) is neutral. Any potential topological defects resulting from the breaking of the underlying symmetry are expected to be diluted away.

The extrema of the potential concerning $\psi$ can be determined by setting $\frac{\partial V}{\partial \psi}=0$, resulting in the following condition,
 \begin{equation}
 \psi^{q-1}=0.
\end{equation}
In addition, we obtain the temporal minimum,
\begin{equation}
 \left(\frac{\psi}{M}\right)^q+\left(\frac{\phi}{\phi_c}\right)^{p}=1.
 \end{equation}
The $q^{\text{th}}$ derivative of the potential is given by,
\begin{equation}
   \frac{1}{\Lambda^4} \frac{\partial^{q}V}{\partial\psi^{q}}\simeq -2\frac{q!}{M^q}+\frac{(2q)!}{q!}\frac{1}{M^q}\left(\frac{\psi}{M}\right)^{q}+2\left(\frac{\phi}{\phi_c}\right)^{p}\frac{q!}{M^q}.
\end{equation}
In the valley region, this expression simplifies to,
\begin{equation}\label{eq39}
\frac{1}{\Lambda^4} \left(\frac{\partial^{q}V}{\partial\psi^{q}}\right)_{\psi=0}\simeq 2\frac{q!}{M^q}\left(\left(\frac{\phi}{\phi_c}\right)^{p}-1\right).
\end{equation}
For even values of $q$, Eq.~(\ref{eq39}) represents a generalized form of the waterfall transition, where the $q^{\text{th}}$ derivative of the potential changes sign around the critical value $\phi_c$. 

In this general case, equations of motion for the scalar fields under slow-roll approximation are,
\begin{equation}\label{N50}
	\phi^\prime(N)\simeq-\frac{p}{\mu}\left(\frac{\phi}{\mu}\right)^{p-1}\left[1+2\left(\frac{\mu}{\phi_c}\right)^{p}\left(\frac{\psi}{M}\right)^{q}\right],
	\end{equation}
 \begin{equation}\label{N51}
	\psi^\prime(N)\simeq\frac{2q}{M}\left(\frac{\psi}{M}\right)^{q-1}\left[\left(1-\left(\frac{\phi}{\phi_{c}}\right)^{p}\right)-\left(\frac{\psi}{M}\right)^{q}\right].
\end{equation}
In the `new inflation limit', the equations of motion (\ref{N50}) and (\ref{N51}) are respectively dominated by second and first terms, i.e.,  $2\left(\frac{\mu}{\phi_c}\right)^{p}\left(\frac{\psi}{M}\right)^{q} \gg 1$ and $\left(1-\left(\frac{\phi}{\phi_{c}}\right)^{p}\right) \gg \left(\frac{\psi}{M}\right)^{q}$.
The equations of motion for $\phi$ and $\psi$ in this `new inflation limit' become,
\begin{align}\label{Gvi}
 \phi^{\prime}(N) &\simeq \frac{-2p}{\phi_{c}}\left(\frac{\phi}{\phi_{c}}\right)^{p-1}\left(\frac{\psi}{M}\right)^{q}, \\
 \psi^{\prime}(N) &\simeq \frac{2q}{M}\left(\frac{\psi}{M}\right)^{q-1}\left(1-\left(\frac{\phi}{\phi_{c}}\right)^{p}\right).
 \label{Gvii}
\end{align}
Combining these two equations, we arrive at the following equation,
\begin{equation}\label{Gix}
\frac{dX}{d\chi} \simeq -\frac{p}{q}\frac{X^{p-1}}{\left(1-X^{p}\right)}\chi.
\end{equation}
In the small $\chi$ limit with $X \simeq 1$, the above equation admits the following approximate solution, 
\begin{equation}\label{Gxiv}
  \phi \simeq \phi_{c}\left(1-\frac{1}{\sqrt{q}}\left(\frac{\psi}{\phi_{c}}\right) \right),
\end{equation}
which reduces to Eq.~(\ref{11}) for $q=2$. 
With $d\phi /  d\psi \simeq - 1 / {\sqrt{q}}$, the above expression implies that
\begin{equation}
    \cos{\theta}\simeq - \frac{1}{\sqrt{1+q}},\qquad \sin{\theta}\simeq \frac{\sqrt{q}}{\sqrt{1+q}}.
\end{equation}
For the adiabatic field defined in Eq.~(\ref{Gxv2}), we arrive at the following relation, 
\begin{equation}
\sigma \simeq \sqrt{\frac{1+q}{q}}\psi.
\end{equation}
Finally, the potential in (\ref{Gi}) can expressed in terms of the adiabatic field $\sigma$ as,
\begin{equation}\label{Gxvi}
	V\left(\sigma\right) \simeq \Lambda^{4}\left[ 1-\frac{\mathcal{A}}{\sqrt{q}} \sigma^{q+1}  \right],
\end{equation}   
where $\mathcal{A}\equiv \frac{4}{M^{q}\phi_{c}\left( \sqrt{\frac{q+1}{q}} \right)^{q+1}}$.
\begin{figure}[t] 
\includegraphics[width=85mm,scale=0.7]{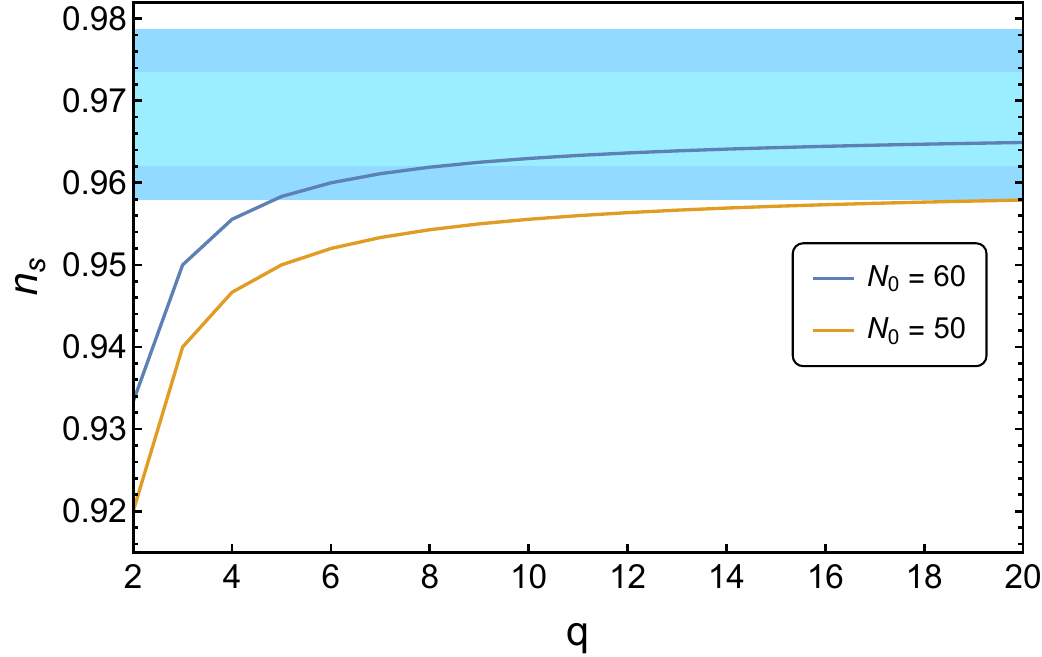}%
\caption{\label{fig:epsart} The scalar spectral index $n_s$ is plotted against the integer power $q$ for $50$ (orange) and $60$ (purple) e-folds. The light (dark) shaded region represents
the Planck 2018 1-$\sigma$ (2-$\sigma$) bounds.}
\end{figure}

\subsection*{The Slow-roll Predictions}
The first and second slow-roll parameters now become,
\begin{equation}\label{i}
    \epsilon\simeq\frac{1}{2}\left( -\frac{\mathcal{A}\left(q+1\right)}{\sqrt{q}}\sigma^{q}\right)^{2}, \qquad \eta \simeq\left( -\mathcal{A}\sqrt{q}\left(q+1\right)\sigma^{q-1}\right).
\end{equation}
The number of e-folds before the end of inflation is,
\begin{equation}\label{iii}
	N_{0} \simeq \int_{\sigma_{e}}^{\sigma_{0}}\left(\frac{V}{V_{\sigma}}\right)d{\sigma} \simeq
 \frac{\sqrt{q}\sigma_{0}^{1-q}}{\mathcal{A}\left(q^2-1\right)},
\end{equation}
where the field value at the end of inflation is given by
\begin{equation}\label{ii}
    \sigma_{e}=\left(\frac{1}{\sqrt{q}\left(q+1\right)\mathcal{A}}\right)^{\frac{1}{q-1}},
\end{equation}
and the field value $\sigma_{0}$ before the last $N_{0}$ number of e-folds is,
\begin{equation}\label{iv}
	\sigma_{0}= \left(\frac{\sqrt{q}}{\mathcal{A}\left(q^{2}-1\right)N_{0}} \right)^{\frac{1}{q-1}}.
\end{equation}
The value of $\Lambda$ can be obtained as
\begin{equation}\label{iv3}
\Lambda  \simeq (24\pi^2 A_s(k_0) \epsilon(\sigma_0))^{1/4}.
\end{equation}

The scalar spectral index can now be expressed in terms of e-folds $N_{0}$ as,
\begin{equation}\label{ns}
    n_{s}  \simeq  1-\frac{2q}{\left(q-1\right)N_{0}},
\end{equation}
with the tensor-to-scalar ratio $r$ given by,
\begin{equation*}\label{vi}
r = 16 \, \epsilon\simeq 16 \left( -\frac{\mathcal{A}\left(q+1\right)}{\sqrt{q}}\left( \frac{\sqrt{q}}{\mathcal{A} \left(q^{2}-1\right)}\frac{1}{N_{0}}\right)^{\frac{q}{q-1}}\right)^{2}.
\end{equation*}
For a typical range of efolds, $N_0 \simeq 50-60$, in the large $q$ limit, we obtain $n_s \simeq 0.96 -0.967$, consistent with the latest Planck bounds, as shown in Fig.~(\ref{fig:epsart}). The predictions of other parameters, for let's say $q = 20$, are as follows: $r \simeq (9-13) \times 10^{-11}$, $\Lambda \simeq 5 \times 10^{-5}$ and $\sigma_0 \simeq 4 \times 10^{-3}$ assuming $\phi_c = M = 0.01$.

\subsection*{Quantum Diffusion Boundary and Initial Conditions} 
\begin{figure}[t!]\centering
\includegraphics[width=0.48\textwidth]{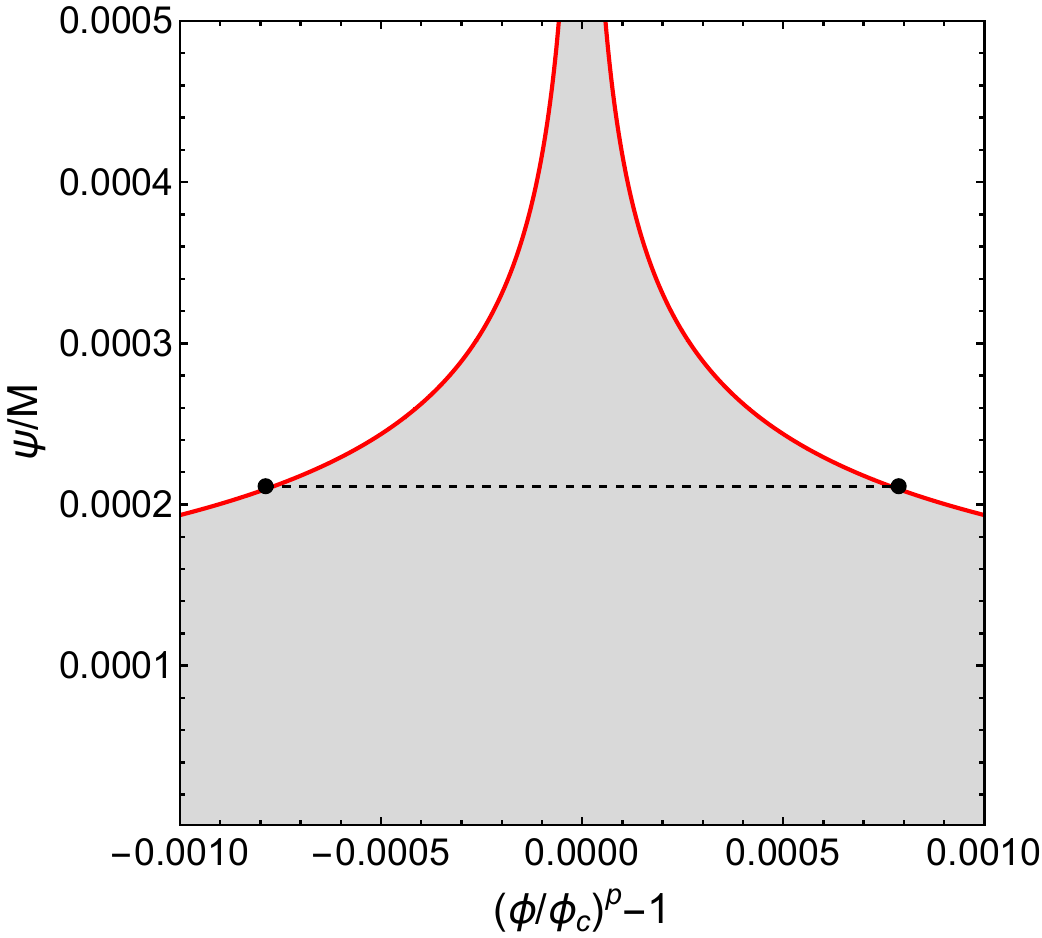}
\caption{\label{fig:DBp4y}
            Quantum diffusion region (gray) for $p=q=4$, where the black line shows the evolution of the fields inside the diffusion region. The $\Lambda$ parameter and the symmetry breaking scale $N$ are respectively $8.87\times 10^{-6}$ and $0.01$ (in Planckian units).}
\end{figure}

In this case, the mass of $\psi$ at $\psi=0$ is zero for $q>2$, i.e., $\frac{\partial^{2}V}{\partial\psi^{2}} = 0$ for $\psi=0$, where $V$ is given in Eq.~(\ref{Gi}). The Langevin equation for $\psi$ in the massless limit leads to, 
\begin{equation}
\frac{d \langle\psi^{2} \rangle}{dN} = \left(\frac{H}{2\pi} \right)^{2}.
\end{equation}
Here, we have ignored the quantum fluctuations in $\phi$. 
In the pre-inflationary phase, when the scalar field $\phi$ is significantly larger than the critical value $ \phi_{c}$, the quantum spread in the field $\psi$ remains nearly zero. However, as $\phi$ approaches the critical instability at $\phi=\phi_{c}$, quantum fluctuations lead to an increase in the value of $ \left\langle \psi^{2} \right \rangle$. 
When the rate of change of the classical field displacement $\delta \psi_{cl}$ over a Hubble time falls below the rate at which quantum fluctuations $\delta \psi_{qu}$ grow over the same time span $H^{-1}$, quantum diffusion becomes important. This condition defines the boundary of the diffusion region as ~\cite{Antusch_2014}, 
\begin{equation}\label{NW3}
  \delta\psi_{qu}=\left(\frac{H}{2\pi} \right) \geq  \delta\psi_{cl} =H^{-1} \left|\dot{\psi} \right|=\left|\frac{\partial_{\psi} V}{V} \right|.
\end{equation}
 Using the generalized potential in Eq.~(\ref{Gi}), we can write, 
\begin{equation}\label{DB01}
	2\,q\left|y^{q-1}\left(1-x^{p} - y^{q}\right) \right| \leq M \frac{  H}{2\pi},
\end{equation}
where $y\equiv \frac{\psi}{M}$ and $x\equiv \frac{\phi}{\phi_{c}}$.
The quantum diffusion boundary defined by Eq.~(\ref{DB01}) is shown in Fig.~\ref{fig:DBp4y} for $q=4$, $M = 0.01$ and $\Lambda \simeq 9 \times 10^{-6}$. The value of $\Lambda$ corresponds to the last 60 numbers of efolds for $\phi < \phi_c$ before the end of inflation. 
The inclusion of this data point serves the sole purpose of elucidating the linkage between the diffusion boundary and the initial conditions, which will become pertinent in our subsequent discussion.
At the diffusion boundary, the value of $y$ exhibits a sharp increase as it approaches $x=1$.
The black curve represents the system's evolution within the diffusion region, bridging the gap between the inflationary phase below $\phi_c$ and its potential initial conditions above $\phi_c$. 
In the following discussion, we will denote the field values at the quantum diffusion boundary as $\psi_{DB}$ and $\phi_{DB}$.

\begin{figure*}[t!]\centering
\subfloat[\label{4a}]
{\includegraphics[width=0.48\textwidth]{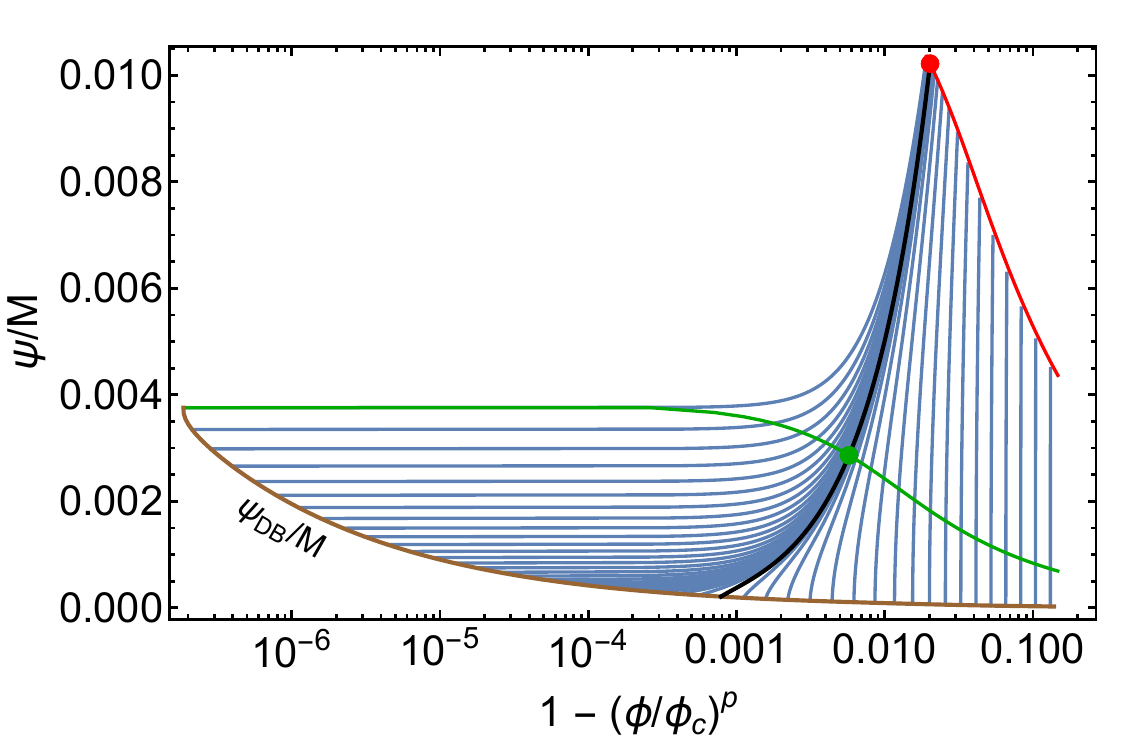}}\qquad
\subfloat[\label{4b}]
{\includegraphics[width=0.45\textwidth]{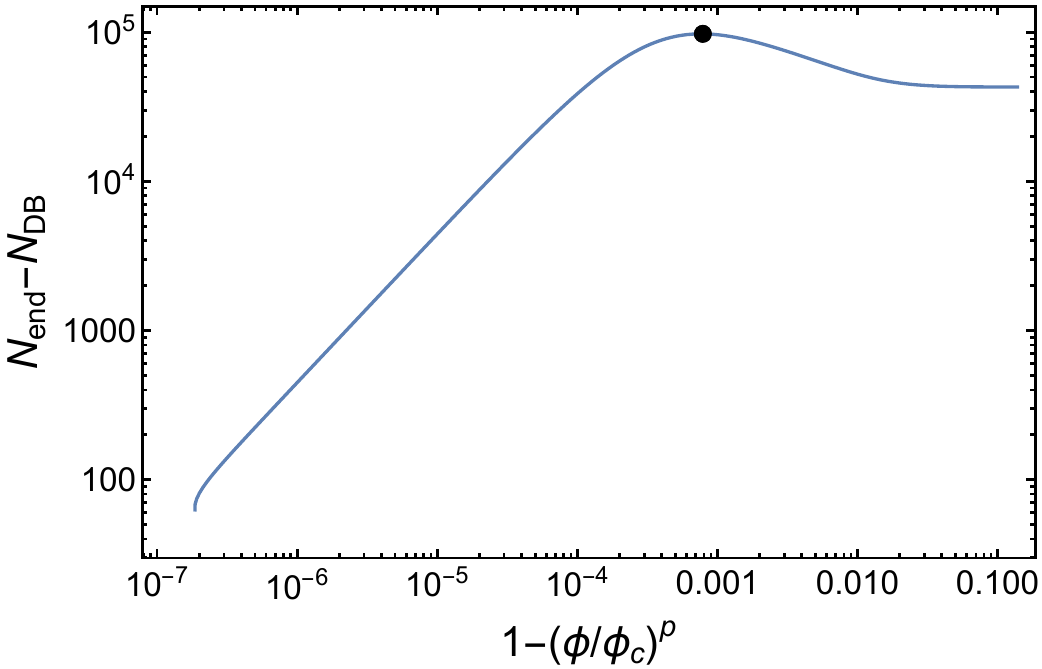}}
\caption{\label{fig:3D}(a)\quad Field trajectories originating from the diffusion boundary (dark red line) for $p=q=4$ are shown. The black curve is the attractor trajectory. The green curve shows the pivot points for the trajectories for the last $60$ efolds with the green (red) dot being the pivot (end) point for the attractor trajectory.\quad(b)\quad Total number of efolds from the diffusion boundary to the end of inflation and the black dot represents the total efolds of the attractor trajectory.}
\end{figure*}

In the diffusion region, the expectation value  $\langle\psi^2\rangle$ grows linearly with efolds and the field value $\psi_{DB}(\phi_{DB} > \phi_c)$ can be related to field value $\psi_{DB}(\phi_{DB} < \phi_c)$  using Eq.~(41) as,
\begin{equation}
\psi_{DB}^2(\phi_{DB} < \phi_c) \simeq \psi_{DB}^2(\phi_{DB} > \phi_c) + \left(\frac{H}{2\pi}\right)^2 \Delta N,
\end{equation}
where $\Delta N$ is the number of efolds that elapsed between the two distinct diffusion boundaries defined by $\phi_{DB} > \phi_c$ and $\phi_{DB} < \phi_c$. The value of $\Delta N$ between these two points can be determined utilizing the slow-roll equation of motion for $\phi$.  
For the data point depicted in Fig.~\ref{fig:DBp4y}, we find that the number of e-folds, $\Delta N$, is approximately on the order of $10^2$. Consequently, we find that $\psi_{DB}^2(\phi_{DB} < \phi_c) \simeq \psi_{DB}^2(\phi_{DB} > \phi_c)$, leading to a linear evolution within the diffusion region. By following this procedure, a connection can be established between the points on both boundaries. Nevertheless, in the subsequent discussion,  we will solely focus on describing the successful points along the diffusion boundary with $\phi_{DB} < \phi_c$.

Ultimately, as the system progresses, it exits the diffusion region when $\phi_{DB} < \phi_{c}$, and classical motion begins to predominate. For our data points, this diffusion boundary resides within phase-2, where the second term in Eq.~(\ref{N50}) takes precedence. It's worth noting that other phases of classical inflation remain confined within the diffusion region and do not bear relevance to the current context. Subsequently, the inflationary phase ends, giving way to an oscillatory period around one of the two global minima:  $\psi=\pm M$ with $\phi=0$.

\subsection*{Inflationary Trajectories from the Diffusion Boundary} 

Next, we study the various inflationary trajectories originating from the diffusion boundary with $\phi_{DB} < \phi_c$. These trajectories are shown in Fig.~\ref{4a} where the diffusion boundary is depicted in dark-red color. The red curve marks the end of inflation whereas the green curve represents the last 60 e-folds before the end of inflation. Note that most of the trajectories are converging towards the black trajectory which will be called an attractor inflationary trajectory. This is the trajectory we have discussed analytically in the previous subsections as the new inflation limit. Note that this trajectory experiences a maximum number of e-folds from the diffusion boundary to the end of inflation as shown in Fig.~\ref{4b}.

In the small $\phi_{DB}$ limit, we obtain a single-field inflation where $\psi$ is playing the role of the inflaton.   In the other extreme where $\phi \simeq \phi_{c}$, both fields participate in the last observable part of the inflation. This limit is multifield inflation that deals with the $\delta N$ formalism for the calculation of its prediction. Beyond this limit, $(\phi/\phi_c)^p-1 < 10^{-7}$,  enough e-folds are not realized and inflation ends quickly. As the number of e-folds in this limit is short we can call this limit the mild waterfall inflation. The attractor inflationary serves as a partition to separate the above-mentioned two extremes. As discussed in the previous subsections, both $\phi$ and $\psi$ fields participate in the attractor inflationary trajectory in such a way that the adiabatic field $\sigma$ experiences no turning and this is an effective single-field inflation model.
For a similar distinction of three different regimes in the waterfall region of tribrid inflation, see ~\cite{Antusch_2014}.

\section*{ The $\delta N $ Formalism}

\begin{figure*}[t!]\centering
\subfloat[\label{fig:6a}]
{\includegraphics[width=0.45\textwidth]{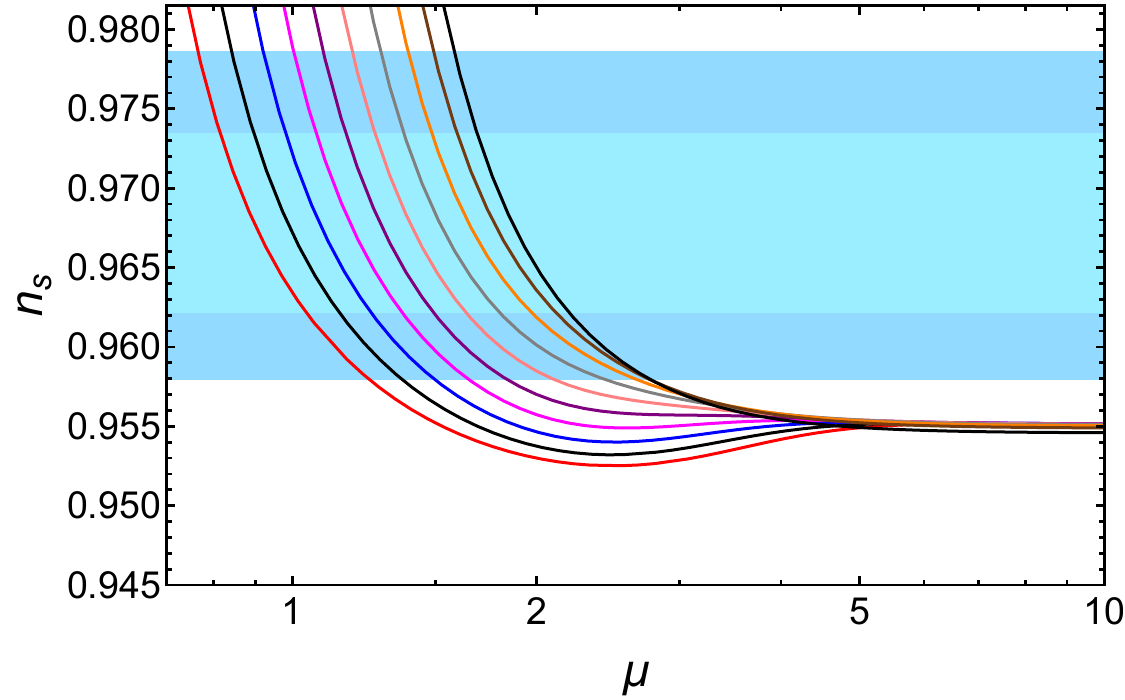}}\qquad
\subfloat[\label{fig:6b}]
{\includegraphics[width=0.45\textwidth]{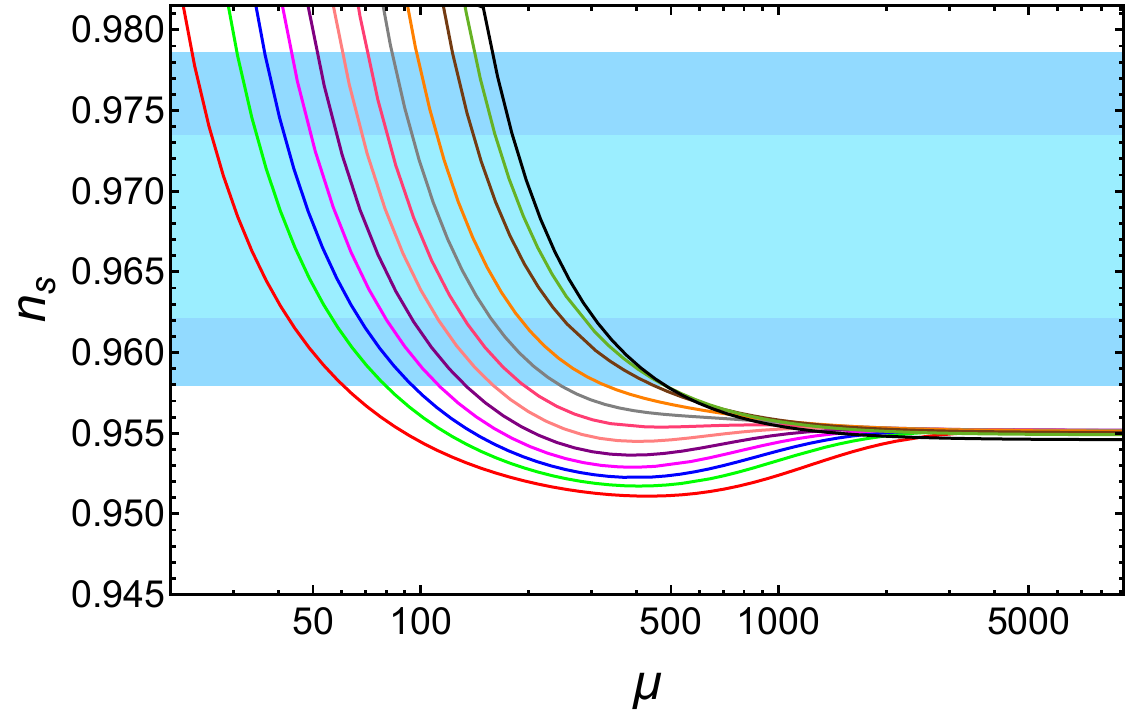}}
\caption{\label{fig:6}The scalar spectral index $n_{s}$ plotted against the mass parameter $\mu$, for (a)\quad $p=4$, $q=4$ for different points ($1-(\phi_{DB}/\phi_{c})^{p}$,$\psi_{DB}/M$)  from  ($2.51189\times10^{-5}, 7.3374\times10^{-4}$) (red) to ($1.99526\times10^{-7},3.67854\times10^{-3}$) (black) and (b)\quad $p=2$, $q=4$ for different points ($1-(\phi_{DB}/\phi_{c})^{p}$,$\psi_{DB}/M$)  from  ($7.94328\times10^{-5}, 5.8393\times10^{-4}$) (red) to ($3.16228\times10^{-7}, 3.68507\times10^{-3}$) (black), for $60$ efolds from the pivot scale. The light (dark) shaded region represents the Planck 2018 1-$\sigma$ (2-$\sigma$) bounds.}
\end{figure*}

The numerical estimates are determined for the power spectrum amplitude of primordial curvature perturbations, $\mathcal{P}$, and the scalar spectral index $n_s$ using the $\delta N$ formalism as given in \cite{Sugiyama_2013}.
The $\delta N$ formalism, which relies on the separate universe approximation, asserts that the curvature perturbation $\zeta(x, t)$ on a spatial hypersurface characterized by a uniform energy density can be expressed as the difference between the number of e-folds realized from an initially flat hypersurface to a final hypersurface with a uniform energy density,
$\delta N^{\mathrm{f}}_{\mathrm{i}}$, 
\begin{equation}\label{dN1}
\zeta = \delta N^{(\mathrm{f})}_{(\mathrm{i})} \equiv N(t,x)-N_{0}(t),
\end{equation}
where we label the initial hypersurface by (i) and the final by (f). In order to extract predictions for the primordial curvature perturbations, our initial hypersurface is chosen at the time $t_{*}$ corresponding to the Hubble exit of the observable pivot scale $k_{*}=0.05$ Mp$c^{-1}$ and the final hypersurface of a constant energy density is chosen at the end of inflation where slow-roll parameter $\epsilon$ reaches unity.

The field perturbations are approximately Gaussian and if the amplitude of these perturbations is very small, then in the slow-roll approximation the curvature perturbations can be expanded as,
\begin{equation}\label{dN2}
\zeta \simeq \sum_{i=1}^{n}N_{i}\delta \phi_{i}^{(\mathrm{i})} +\frac{1}{2} \sum_{i,j=1}^{n} N_{ij}\delta \phi_{i}^{(\mathrm{i})}\phi_{j}^{(\mathrm{i})},
\end{equation}
where $N_{i}=\frac{\partial{\delta N_{(\mathrm{i})}^{(\mathrm{f})}}}{\partial\phi_{i}^{(\mathrm{i})}}$ and 
$N_{ij}=\frac{\partial^{2}{\delta N_{(\mathrm{i})}^{(\mathrm{f})}}}{\partial\phi_{i}^{(\mathrm{i})} \partial \phi_{j}^{(\mathrm{i})}}$.
The power spectrum amplitude and the scalar spectral index can be calculated as, 
\begin{align} \label{dN3}
 \mathcal{P}_{\zeta}(k_{*})   &= \frac{H^{2}_{*}}{4\pi^{2}}\sum_{i=1}^{n}N_{i}^{2} ,    \\
   n_{s}   &= 1-2 \epsilon_* + \frac{\Sigma_{ij}{\dot{\phi_{i}}}_{*}N_{j}N_{ij}}{H_{*}\Sigma_{i}N^{2}_{i}}, 
\end{align}
where $\epsilon = -\frac{\dot{H}}{H^{2}}$ is the slow-roll parameter. In the effective singlet field scenario, $\mathcal{P}_{\zeta}(k_{*})$ reduces to $A_s(k_*)$ defined in Eq.~(\ref{iv3}). The non-Gaussianity of the perturbation is parametrized by $f_{NL}$, which is given by
\begin{equation}
f_{NL}   = \frac{5}{6}\frac{\Sigma_{ij}{N_{i}}N_{j}N_{ij}}{\left( \Sigma_{i}N^{2}_{i} \right)^{2}}.
\end{equation}

\begin{figure*}[t!]\centering
\subfloat[\label{7a}]
{\includegraphics[width=0.415\textwidth,height=5cm]{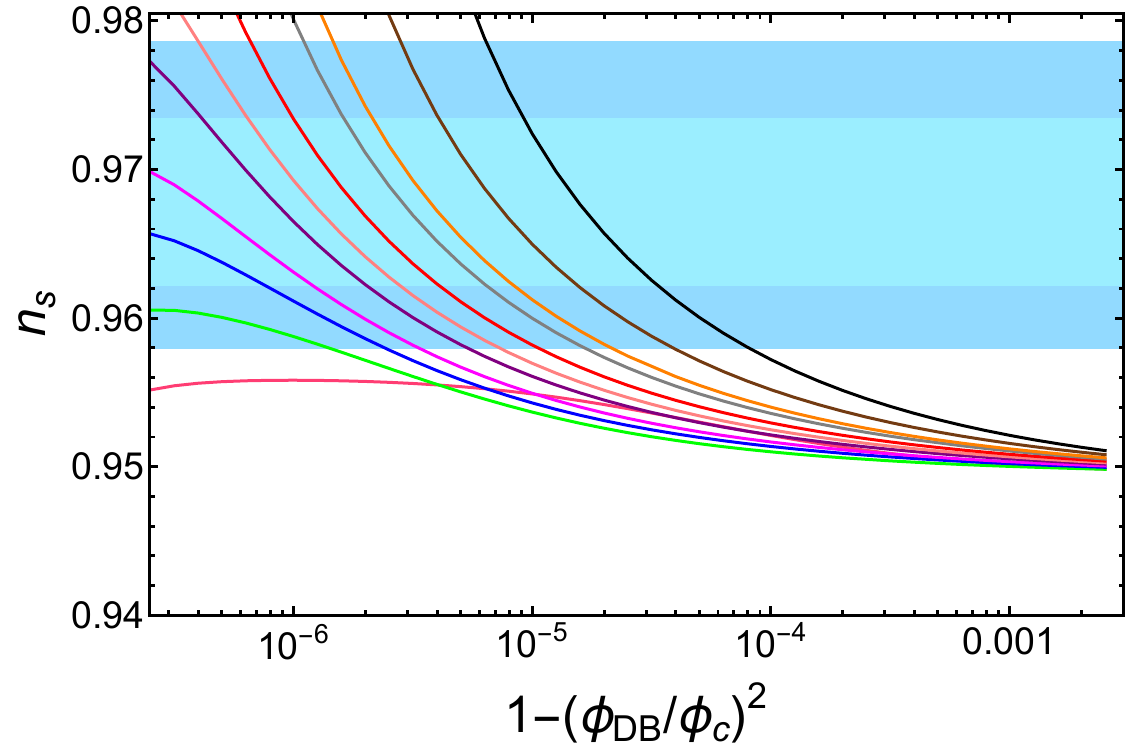}}\qquad
\subfloat[\label{7b}]
{\includegraphics[width=0.52\textwidth,height=5cm]{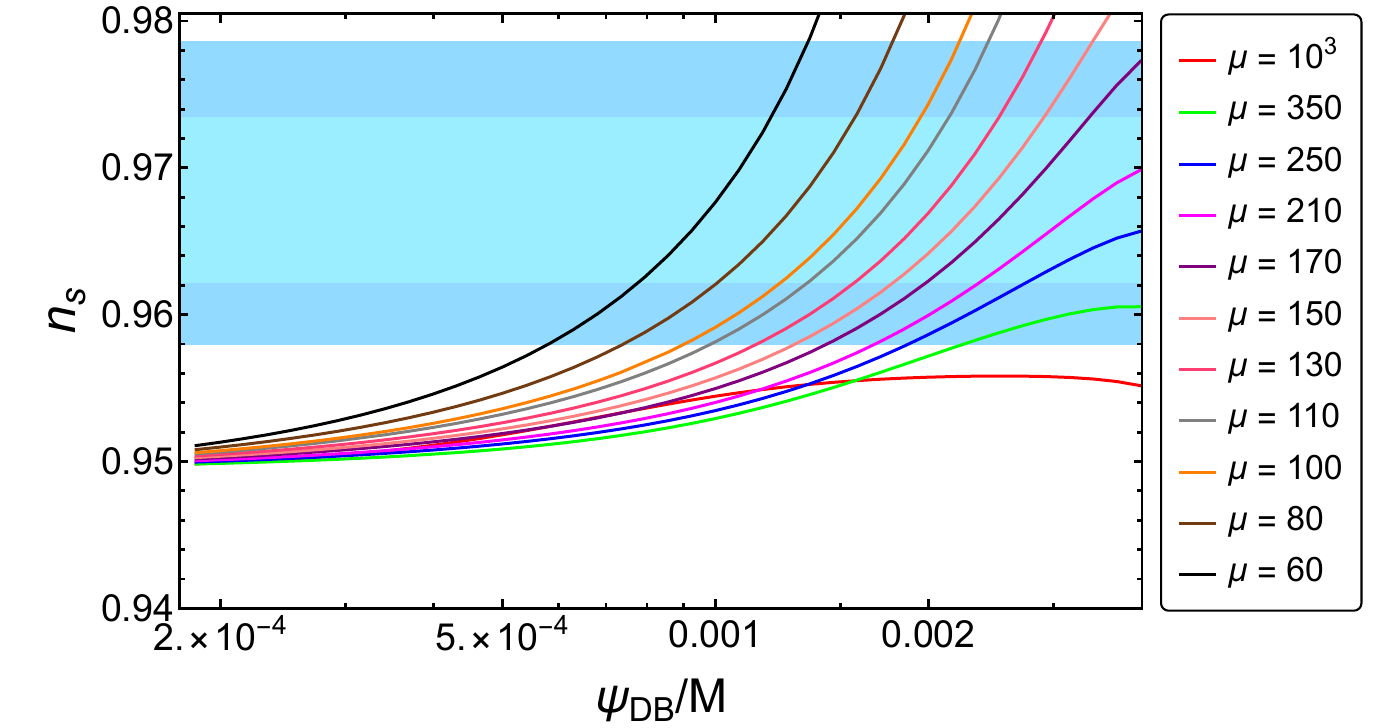}}\qquad
\subfloat[\label{7c}]
{\includegraphics[width=0.415\textwidth,height=5cm]{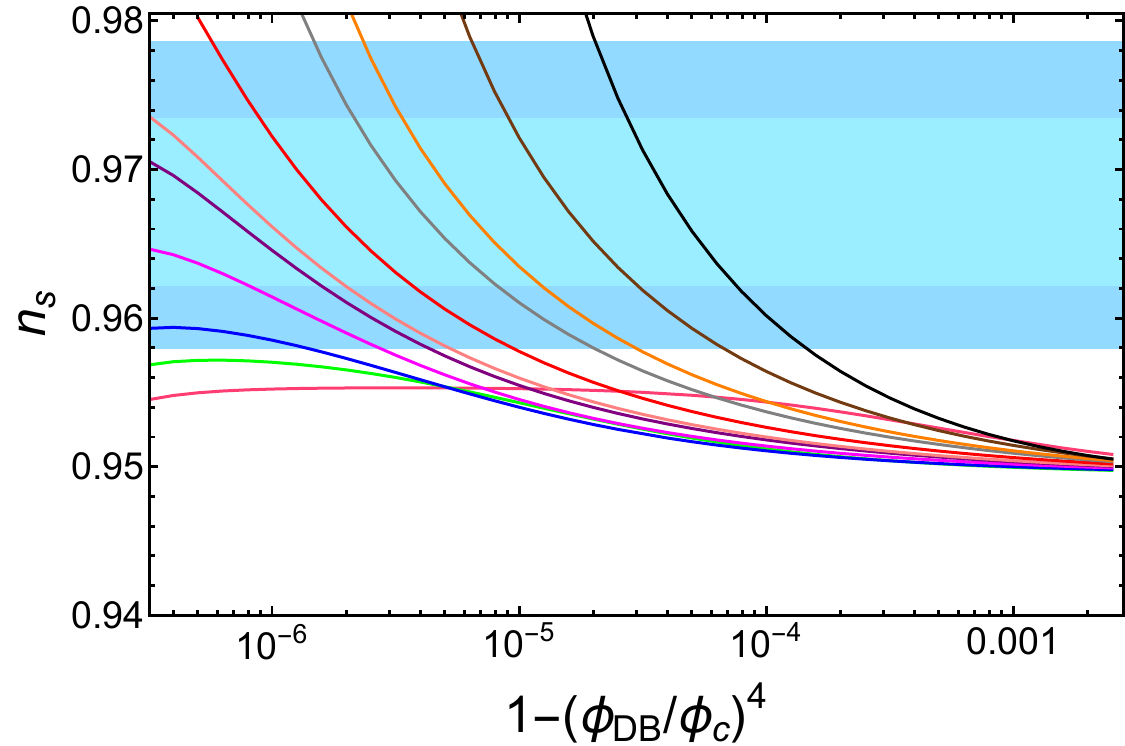}}\qquad
\subfloat[\label{7d}]
{\includegraphics[width=0.52\textwidth,height=5cm]{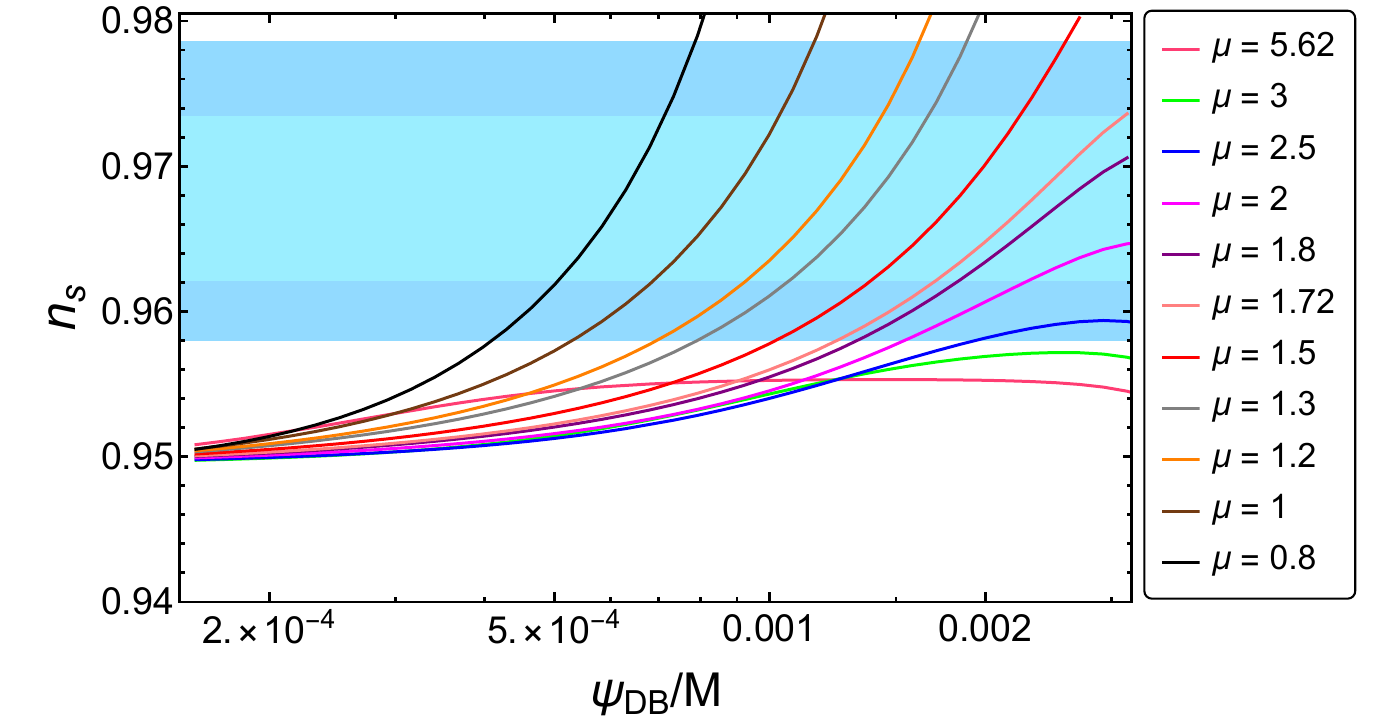}}
\caption{\label{fig:7}Top Panels: Plot of the scalar spectral index $n_{s}$ against (a)\quad$1-(\phi_{DB}/\phi_c)^p$, \quad(b)\quad$\psi_{DB}/M$ over the diffusion boundary, using different $\mu$ values, for $p=2$ \& $q=4$ and  $60$ efolds before the end of inflation.  Bottom Panels:  Plot of the scalar spectral index $n_{s}$ against (c)\quad$1-(\phi_{DB}/\phi_c)^p$, \quad(d)\quad$\psi_{DB}/M$ over the diffusion boundary, using different $\mu$ values, for $p=4$ \& $q=4$ and  $60$ efolds before the end of inflation. The light (dark) shaded region represents the Planck 2018 1-$\sigma$ (2-$\sigma$) bounds.}
\end{figure*} 

\subsection*{Numerical Results for Inflationary 
Trajectories from the Diffusion Boundary}
We have conducted numerical work to calculate inflationary trajectories emerging from the diffusion boundary for a wide range of values of the parameter $\mu$, using the $\delta N$ method. The results are depicted in Fig.~\ref{fig:6} for the scalar spectral index as a function of $\mu$ and in Fig.~\ref{fig:7} for field values along the diffusion boundary. Specifically, we consider cases where $(p,\,q) =(2,4)$  or $(p,\,q) = (4,4)$.
It is worth noting that for smaller values of the mass parameter $\mu$, these initial conditions can yield inflationary predictions consistent with the most recent Planck data.
We maintain a fixed symmetry-breaking scale, $M\simeq 2\times 10^{16}$~GeV, throughout our numerical analysis. Additionally, we adjust the inflationary scale parameter $\Lambda$ to ensure that it matches the measured amplitude of scalar perturbations.

It's important to note that successful inflation can be achieved even with smaller values of the parameter $q$, specifically $q\gtrsim 4$. We explore two scenarios, denoted as $(p,,q) = (2,4)$ and $(p,,q) =(4,4)$, where the value of $q$ remains the same while $p$ varies. This choice allows us to demonstrate that the outcomes are primarily influenced by the value of $q$.
To ensure that our predictions for the scalar spectral index $n_{s}$ align with the Planck bound at the $1-\sigma$ level, we find that the mass parameter $\mu$ falls within the following ranges:
\begin{equation}\label{eq50}
\begin{aligned}
  20 \lesssim \mu   \lesssim 2500 , \text{ for } (p,\,q) = (2,4),\\[18pt]
   1  \lesssim \mu   \lesssim 5 , \text{ for } (p,\,q) = (4,4).
  \end{aligned}
\end{equation}

Interestingly, the shape of the results is consistent between these two cases, with the primary difference being a rescaling of the $\mu$ parameter. This relationship can be understood as follows: for the same value of $q$, the $\mu$ parameter corresponding to two different values of $p$, denoted as $p_1$ and $p_2$, can be related as:
\begin{equation}\label{mup}
\left(\frac{\mu_{p_1}}{\phi_c} \right)^{p_1} \simeq \left(\frac{\mu_{p_2}}{\phi_c} \right)^{p_2}.
\end{equation}
Here, $\mu_1$ ($\mu_2$) represents the value of $\mu$ associated with $p_1$ ($p_2$). Consequently, for the case where $p_1 = 2$ and $p_2 = 4$, we find that $\mu_1 \simeq 100 \, \mu_2^2$. This relationship explains the connection between the two ranges quoted in the above Eq.~(\ref{eq50}) for different values of $p$.

When $\mu$ takes values smaller than those specified in Eq.~(\ref{eq50}), we observe that the scalar spectral index increases toward unity, deviating from Planck's bounds. This behavior is clearly depicted in Fig.~\ref{fig:6a} and Fig.~\ref{fig:6b}.
It's noteworthy that the realistic range of $\mu$ outlined in Eq.~(\ref{eq50}) corresponds to phase-1, characterized by the condition $2(\mu/\phi_c)^p (\psi_{DB}/M)^q < 1$. This contrasts with the `new inflation limit' featuring an attractor solution, which falls within phase-2 and is associated with larger values of $\mu$.
In the cases we have investigated, the `new inflation limit' featuring an attractor solution is realized for significantly larger values of $\mu$:
\begin{equation*}
\mu \gg 2500 ,  \text{ for } (p,\,q) = (2,4),
\end{equation*}
and
\begin{equation*}
  \mu  \gg  5 , \text{ for } (p,\,q) = (4,4).
\end{equation*}
In this limit, as depicted in Fig.~\ref{fig:7}, we do not achieve the central value of $n_{s}$ for the entire field range spanning the diffusion boundary. 
This aligns with our previously obtained findings for the case with $q=4$, as illustrated in Fig.~\ref{fig:epsart}. Nevertheless, for larger values of $q$, we can obtain scalar spectral index values consistent with Planck data at the $1\sigma$ level, as previously demonstrated in Fig.~\ref{fig:epsart}. 
Finally, it's worth noting that the non-gaussianity parameter, denoted as $f_{NL}$, is found to be slow-roll suppressed, having a value on the order of  $10^{-2}$, which is well within the current bounds set by Planck \cite{Plank_2018}.

\subsection*{Beyond Minimal Generalization} 

We should make a few comments regarding a potential extension of the minimal generalization we discussed earlier for the potential described by Eq.~(\ref{Gi}). Let's consider the following potential generalization:
\begin{equation} \label{Gi1}
\small V = \Lambda^4 \left[\left(1-\left(\frac{\psi}{ M}\right)^{q_1}\right)^2 +\left(\frac{\phi}{\mu}\right)^{p_1} + 2 \left(\frac{\phi}{\phi_{c}}\right)^{p_2} \left(\frac{\psi}{M}\right)^{q_2} \right].
\end{equation}
Note that the critical value $\phi_c$ in the above expression describes the waterfall transition point only for $q_1=q_2$. In the case where the value of the $\phi$ field is dynamically suppressed before observable inflation begins, the above potential simplifies to:
\begin{equation} \label{Gi2}
V \simeq \Lambda^4 \left(1 - 2 \left(\frac{\psi}{ M}\right)^{q_1} \right).
\end{equation}
This situation allows for a new inflation scenario. For $q_1 \neq q_2$, the potential in Eq.~(\ref{Gi1}) doesn't exhibit a waterfall. For $q_1 < q_2$, it rather supports a smooth hybrid inflation mechanism \cite{Lazarides_1995,Lazarides_1996,Yamaguchi_2004,_eno_uz_2004,Rehman_2012,Rehman:2014rpa}. For $q_1 > q_2$, the classical evolution of $\phi$ in the $\psi=0$ valley towards the origin can provide a significant suppression of the term $(\phi / \phi_c)^{p_2}$ for new inflation to work smoothly.

Now, let's consider a mixed scenario in the context of a tribrid inflation model, where the scalar potential can be expressed as:
\begin{eqnarray} \label{G2}
V &=&  \Lambda^4 \left[\left(1-\left(\frac{\psi}{ M}\right)^{q_1}\right)^2 +\left(\frac{\phi}{\mu}\right)^{p} \right. \nonumber \\
&+&  \left.  2 \left(\frac{\phi}{\phi_{c}}\right)^{q_2} \left(\frac{\psi}{M}\right)^{q_1} + 2 \left(\frac{\phi}{\phi_{c}}\right)^{q_1} \left(\frac{\psi}{M}\right)^{q_2} \right].
\end{eqnarray}
If the first mixing term dominates, with $q_1 < q_2$, the new inflation scenario proceeds similarly to the minimal generalized case discussed earlier. However, if the second mixing term dominates, with $q_1 > q_2$, it's necessary to adequately suppress the term $(\phi / \phi_c)^{q_2}$ for new inflation to work effectively. This mixed scenario is studied in ~\cite{Antusch_2014}, where they consider specific values such as $q_1=4, q_2=2$, along with additional mass terms for both fields induced by supergravity corrections. Within the supergravity framework, incorporating correction terms for the $\psi$ field opens up the potential for achieving a waterfall transition, even when $q_1 \neq q_2$, as discussed in \cite{Antusch_2005,Antusch_2014}. Having outlined these potential generalizations beyond the minimal version, we will now focus exclusively on the minimal generalization in the rest of this paper.
\section{Supersymmetric Realization} \label{sec-4}
We begin with a superpotential that exhibits a generalized tribrid structure and can be expressed as follows:
\begin{equation}\label{I1}
W=\kappa\left[S\left(\frac{\Psi^q}{M_c^{q-2}} - M_*^2\right) +\lambda\frac{\Psi^{m}\Phi^n}{M_c^{m+n-3}} \right].
\end{equation}
Here, the integers $m, n$ and $q$ will be specified later, while $\kappa$ and $\lambda$ represent dimensionless couplings. Additionally, $M_c$ denotes the cutoff scale, $S$ is a gauge singlet chiral superfield, and $\Psi$ and $\Phi$ could be gauge non-singlet chiral superfields, depending on the specific model under consideration. 
For instance, in the models considered in \cite{Masoud_2021, Masoud:2021prr}, $\Psi$ corresponds to the GUT Higgs superfield, and $\Phi$ represents the gauge singlet or non-singlet sneutrino. Furthermore, $\Phi$ might represent a gauge-invariant combination of matter and Higgs superfields. The general form of $W$ has already been employed in \cite{Antusch:2012bp} for pseudosmooth tribrid inflation and in \cite{Antusch:2012jc} for K\"{a}hler-driven Tribrid Inflation. The global SUSY minimum of $W$ is characterized by,
\begin{equation}\label{I1a}
\langle S \rangle=0, \quad \langle \Phi \rangle=M \text{ with }
M^q = M_c^{q-2}M_*^2,
\end{equation}
where $M$ is the symmetry-breaking scale. To facilitate future discussions, the mass parameter $M_*$ is eliminated in terms of $M$ and $M_c$ using the relationship in the above equation.
The tribrid form of the superpotential is chosen because its scalar potential can reduced to the generalized potential form presented in Eq.~(\ref{Gi}). To facilitate subsequent analysis, we absorb the parameter $\kappa$ into a rescaling of the scalar potential, and all remaining parameters are assumed to be real.

The form of the above superpotential can be restricted by the $U(1)_R \times Z_q$ symmetries. In Table \ref{tab:table1}, we outline the charge assignments under these symmetries for the various superfields. Especially, the $Z_q$ charge assigned to the last term is $ m + n = q$. The global SUSY F-term scalar potential corresponding to the superpotential $W$ is expressed as,
\begin{equation}\label{I2}
\begin{split}
    V &=\left|\frac{\Psi^{q} - M^{q} }{M_c^{q-2}}\right|^{2}
    + \left|q\frac{S\Psi^{q-1}}{M_c^{q-2}} +m\lambda \frac{\Psi^{m-1}\Phi^n}{M_c^{m+n-3}}\right|^{2}\\&
    +\left|n\lambda\frac{\Psi^{m}\Phi^{n-1}}{M_c^{m+n-3}}\right|^{2}.
\end{split}
\end{equation}
Note that we use the same notation for the scalar components of the chiral superfields $S$, $\Phi$, and $\Psi$ here.

When $\Phi$ is set to zero, the above potential simplifies to a generalized form of the hybrid inflation potential. However, this does not yield the desired form of the scalar potential (\ref{Gi}). On the other hand, if we consider the scenario where $S$ vanishes, the scalar potential adopts the form of the generalized scalar potential in (\ref{Gi}), provided we choose $m=\frac{q+2}{2}$ and $n=\frac{q-2}{2}$. This identification helps clarify our rationale for selecting supersymmetric tribrid inflation as the framework for investigating the new inflation phase in the waterfall region.

Let's delve into the details of the setup further. We begin with the K\"{a}hler potential expressed as follows:
\begin{equation}\label{k}
\begin{split}
K&=|S|^{2}+|\Psi|^{2}+|\Phi|^{2}+\kappa_{S} \frac{|S|^{2}}{4}+\kappa_{\Psi} \frac{|\Psi|^{2}}{4}+\kappa_{\Phi} \frac{|\Phi|^{2}}{4}\\&
+\kappa_{S\Psi} |S|^{2}|\Psi|^{2}
			+\kappa_{S\Phi} |S|^{2}|\Phi|^{2}+\dots
\end{split}
\end{equation}
 \begin{table}[t]
\caption{\label{tab:table1}%
Charge assignments of the superfields under $U(1)_R \times Z_q$.
}
\begin{ruledtabular}
\begin{tabular}{lcdr}
\textrm{Superfield}&
\textrm{$U(1)_R$}&
\multicolumn{1}{c}{\textrm{$Z_q$}}\\
\colrule
S & 1 & 0\\
$\Psi$ & 0 & 1\\
$\Phi$ & 1/$n$ & 1\\
\end{tabular}
\end{ruledtabular}
\end{table}
In the context of this setup, the F-term supergravity (SUGRA) scalar potential is given by,
\begin{equation}\label{k1}
    V=e^{K}\left(
    K_{i\bar{j}}^{-1}D_{z_{i}}WD_{z^{*}_j}W^{*}-3 \left| W\right| ^{2}\right) + V_{D},
\end{equation}
with,
\begin{equation}\label{k2}
    D_{z_{i}}W \equiv \frac{\partial W}{\partial z_{i}}+\frac{%
			\partial K}{\partial z_{i}}W , \,\,\,
    K_{i\bar{j}} \equiv \frac{\partial ^{2}K}{\partial z_{i}\partial z_{j}^{*}},
\end{equation}
and 
\begin{equation}\label{k3}
	D_{z_{i}^{*}}W^{*}=\left( D_{z_{i}}W\right)^{*},
\end{equation}
with $z_{i}$ being the bosonic components of the superfields $z_{i}\in \{S,\Psi,\Phi,\cdots\}$ . Since the scalar fields in this setup are complex, they can be represented as
\begin{equation}\label{I3}
    \Psi = \psi \, e^{i\theta_{\psi}}, \, S = \left|S\right| e^{i\theta_{S}}, \, \Phi = \phi \, 
 e^{i\theta_{\phi}}.
\end{equation}
We further assume that the phases of these fields are stabilized at their respective minima, 
\begin{equation}
\theta_{\psi} =0,  \quad  \theta_{S} + n \, \theta_{\phi}=0.
\end{equation}
To understand how these phases might influence inflationary dynamics, refer to ref.~\cite{Nolde_2013}. Consequently, the potential simplifies to
\begin{equation}\label{I7}
\begin{split}
    V & = V_0 \left(1-\left(\frac{\psi}{M}\right)^q\right)^2 \\ & + \left(q\frac{S \psi^{q-1}}{M_c^{q-2}} +m\lambda\frac{\psi^{m-1}\phi^n}{M_c^{m+n-3}}\right)^{2}
   + n^{2}\lambda^{2}\frac{\psi^{2m}\phi^{2\left(n-1\right)}}{M_c^{m+n-3}} \\ 
    & + 3\kappa_{S}~H^2 S^{2}  + V_0( \alpha~\phi^{2} - \beta~\psi^{2}). 
\end{split}
\end{equation}
with
\begin{equation}
V_0 = \kappa^2 M_c^4  \left( \frac{ M }{ M_c} \right)^{2q} .
\end{equation}
Here,  $\alpha=\kappa_{S\Phi}-1 $, $\beta=\kappa_{S\Psi}-1 $, and $H^{2}\simeq V_0 / 3$ is the Hubble parameter. It's worth noting that a SUGRA mass on the order of $H$ is necessary for the $S$ field to suppress its VEV during inflation.
\begin{figure*}[t!]\centering
\subfloat[\label{fig:8a}]
{\includegraphics[width=0.45\textwidth]{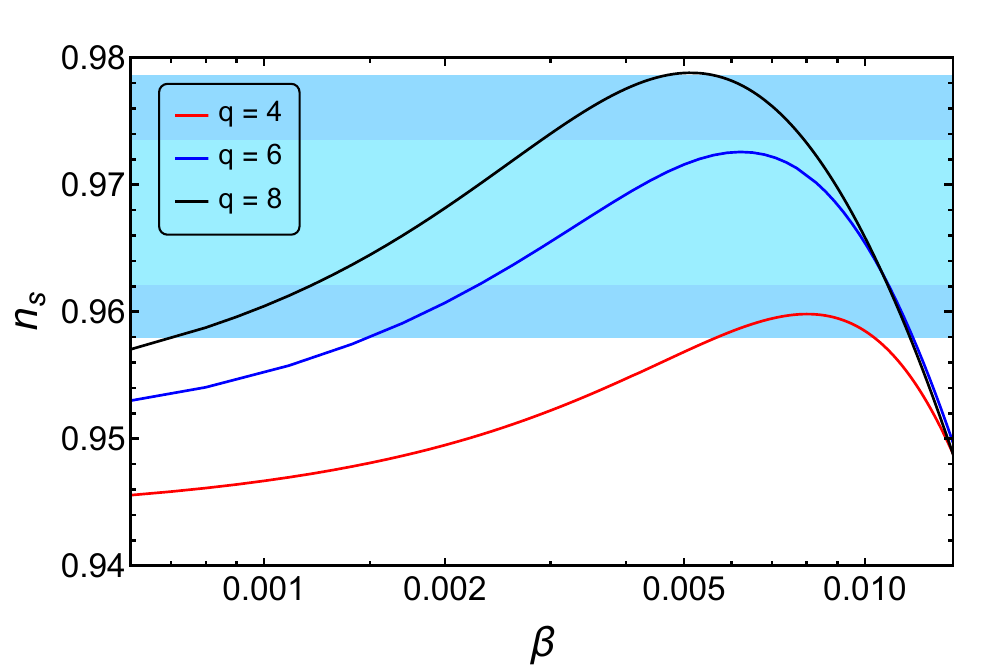}}\quad
\subfloat[\label{fig:8b}]
{\includegraphics[width=0.45\textwidth]{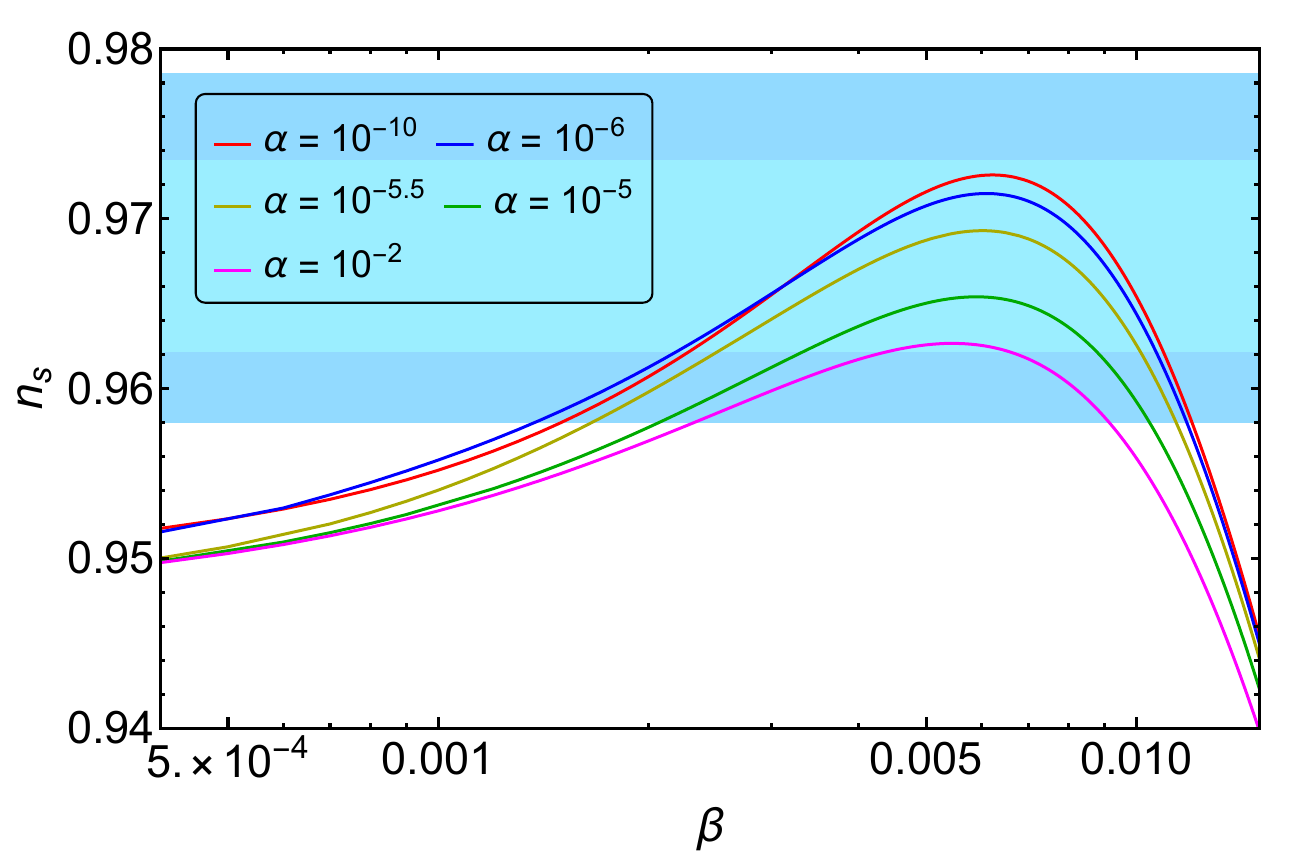}}
\caption{\label{fig:8}(a)\quad The scalar spectral index, $n_s$, plotted against $\beta$ for ($q = 4, \, 6, \, 8$) corresponding to ($n=1,\, 2,\, 3$). The parameter $\alpha$ is approximately zero. (b)\quad The scalar spectral index for $p=6$, plotted against $\beta$ for different values of $\alpha$. The boundary point is fixed in both (a) and (b).  The light (dark) shaded region represents
the Planck 2018 1-$\sigma$ (2-$\sigma$) bounds.}
\end{figure*}
We determine the VEV, $\langle S\rangle $, by minimizing the potential, i.e. $\frac{\partial V}{\partial S}=0 $. The result is
\begin{equation}\label{I9}
\begin{split}
	\langle S \rangle = - \frac{mq \lambda \psi^{m+q-2} \phi^n M_c^{2}} {- q^{2}\psi^{2q-2}/2 +3\kappa_{S}H^{2}}. 
 \end{split}
\end{equation}
In the preinflation phase where $\psi=0$, the driving field $S$  with the Hubble size mass $\left(\kappa_{S} \lesssim -\frac{1}{3}\right)$, quickly settles to its minimum at $\langle S \rangle =0$. Consequently, the potential becomes,
\begin{equation}\label{114}
\begin{split}
V \simeq V_0 &\left[\left(1-\left(\frac{\psi}{M}\right)^q\right)^2+ \lambda^2\left(\frac{q+2}{2}\right)^2 \frac{M_c^2}{M^2} \frac{\phi^{q-2}}{M^{q-2}} \frac{\psi^q}{M^q} \right.\\&
\left.-\beta~\psi^2+\alpha~\phi^{2}\Biggr.\right],
\end{split}
\end{equation}
where we retain the term with a lower power in $\psi$ since the term with a higher power in $\psi$ is relatively suppressed.

Next, we aim to find the critical value of $\phi$ that marks the transition to the waterfall phase. To do this, we calculate the qth derivatives of the above potential, yielding 
\begin{equation}\label{I15}
\small \begin{split}
	 \frac{\partial^{q}V}{\partial\psi^{q}}\simeq V_0 \left[-\frac{2(q!)}{M^q}+\frac{(2q!)}{q!}\psi^q + q!\left(\frac{q+2}{2}\right)^2 \lambda^2 \frac{M_c^2}{M^{q+2}} \frac{\phi^{q-2}}{M^{q-2}}
 \right].
\end{split}
\end{equation}
It is worth noting that the $\beta$-term only contributes when $q = 2$, but it remains negligibly small within the relevant $\beta$ range (less than $10^{-2}$).
Evaluating this derivative at $\psi=0$, we get
\begin{equation}\label{I15}
\begin{split}
   \left. \frac{\partial^{q}V}{\partial\psi^{q}}\right|_{\psi=0}&\simeq V_0 \left[-\frac{2(q!)}{M^q} +
   q!\left(\frac{q+2}{2}\right)^2 \lambda^2 \frac{M_c^2}{M^{q+2}} \frac{\phi^{q-2}}{M^{q-2}} \right].
\end{split}
\end{equation}
The waterfall transition occurs when,
\begin{equation}
	\left(\frac{\partial^{q}V}{\partial\psi^{q}}\right)_{\psi=0,\phi= \phi_{c}}=0.
\end{equation}
Therefore, the critical value of $\phi$ where the transition takes place can be expressed as
\begin{equation}\label{I17}
   \frac{\phi_c^{q-2}}{M^{q-2}}  = \frac{2}{\lambda^{2} \left(\frac{q+2}{2}\right)^{2}} \frac{M^2}{M_c^2}.
\end{equation}
Utilizing this equation, the normalized scalar potential adopts the following form,
\begin{equation}\label{I20}
\begin{split}
    V/V_{0}&\simeq \left(1-\left(\frac{\psi}{M}\right)^{q}\right)^{2}+2\left(\frac{\psi}{M}\right)^{q}\left(\frac{\phi}{\phi_{c}}\right)^{2n}\\&
     +\alpha~\phi^{2} - \beta~\psi^{2},
\end{split}
\end{equation}
where $2n = q-2$. With canonically normalized fields, we perform the replacements, $\psi \rightarrow\frac{\psi}{\sqrt{2}}$ and $\rightarrow\frac{\phi}{\sqrt{2}}$.
This adjustment brings the above potential in line with the required generalized form of Eq.~(\ref{Gi})  with $2n = q-2 = p$, $\beta=0$, $\alpha (p=2) \simeq \mu^{-2}$, supplemented with the replacements, $M \rightarrow\frac{M}{\sqrt{2}}$ and $\phi_c \rightarrow\frac{\phi_c}{\sqrt{2}}$ (with $M_c \rightarrow\frac{M_c}{\sqrt{2}}$).
However, there are still some differences due to the presence of $\alpha$ and $\beta$ terms. These differences render the predictions of this specific supersymmetric realization particularly intriguing, as discussed in the subsequent section.

\subsection*{Results and Discussion} 
In the current supersymmetric framework, the diffusion boundary is given by,
\begin{equation}
	2 M^2\beta~y+2\,q\left|y^{q-1}\left(1-x^{2n} - y^{q}\right) \right| \leq M \frac{  H}{2\pi}.
\end{equation}
For $ M^2 \beta y \gg  q y^{q-1}$, the above condition simplifies to, 
\begin{equation}
	 \psi \leq  \psi_{DB} \equiv \frac{H}{4\pi \beta}.
\end{equation}
For $\beta \simeq 10^{-3}-10^{-2}$, $M = 0.01$, and $H \simeq 10^{-10}$, we obtain $\psi_{DB} \simeq 10^{-9}-10^{-8}$. In order for classical evolution to dominate, it in necessary for $\psi$ to exceed $\psi_{DB}$.

In the limit of large values for $q$ and small values for $\alpha$ and $\beta$, the potential in Eq.~\ref{I20} predicts the scalar spectral index in accordance with Planck data, Therefore, in this section, we focus on the smallest possible values of $q$ and explore the ranges of $\alpha$ and $\beta$ that remain consistent with the observational data.

In Fig.~\ref{fig:8a}, we present a plot depicting the scalar spectral index as a function of the parameter $\beta$ for three different values of $n$,  (specifically, $n =1,\,2$ and $n=3$, which correspond to $q=4,\,6$ and $8$, respectively) while assuming a negligible value for $\alpha$ (i.e., $\alpha \ll 1$). For $q=6$ and $q=8$, the range $\beta \simeq 0.001-0.01 $ falls within the 1-$\sigma$ Planck constraint on $n_s$, whereas $q=4$ is just within the 2-$\sigma$ range around $\beta \sim 0.008$.
It's worth noting that the smallest possible value for $n$, which is $n = 1$, leads to a high supersymmetry breaking scale, making it less relevant for physics related to the Large Hadron Collider (LHC). Therefore, for our subsequent discussions, we adopt $n=2$ (or $q=6$) as the next smallest value, which also remains consistent with the Planck constraints over a wider range of $\beta$.

\begin{figure*}[t!]\centering
\subfloat[\label{fig:9a}]
{\includegraphics[width=0.466\textwidth,height=0.295\textwidth]{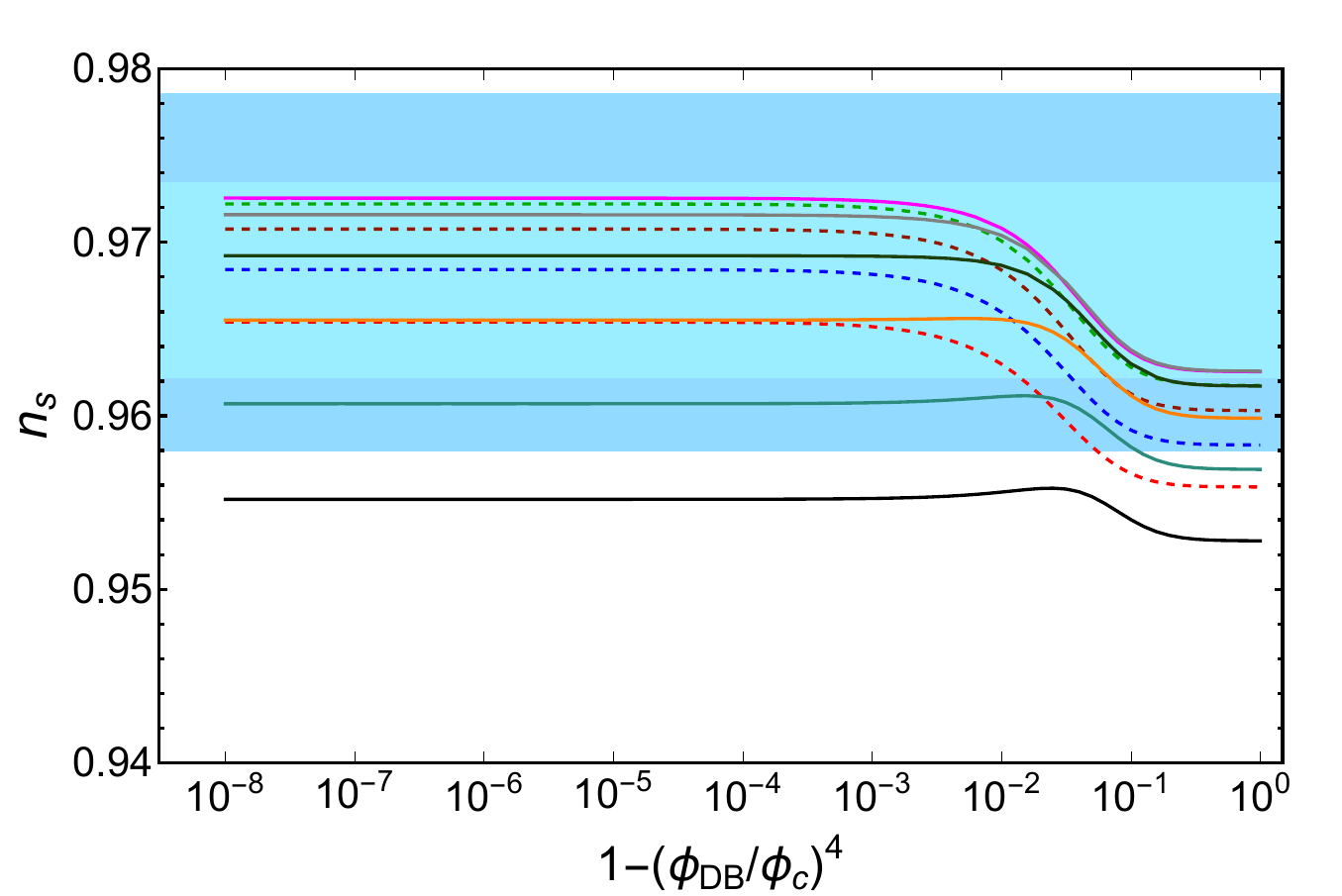}}
\subfloat[\label{fig:9b}]
{\includegraphics[width=0.551\textwidth,height=0.295\textwidth]{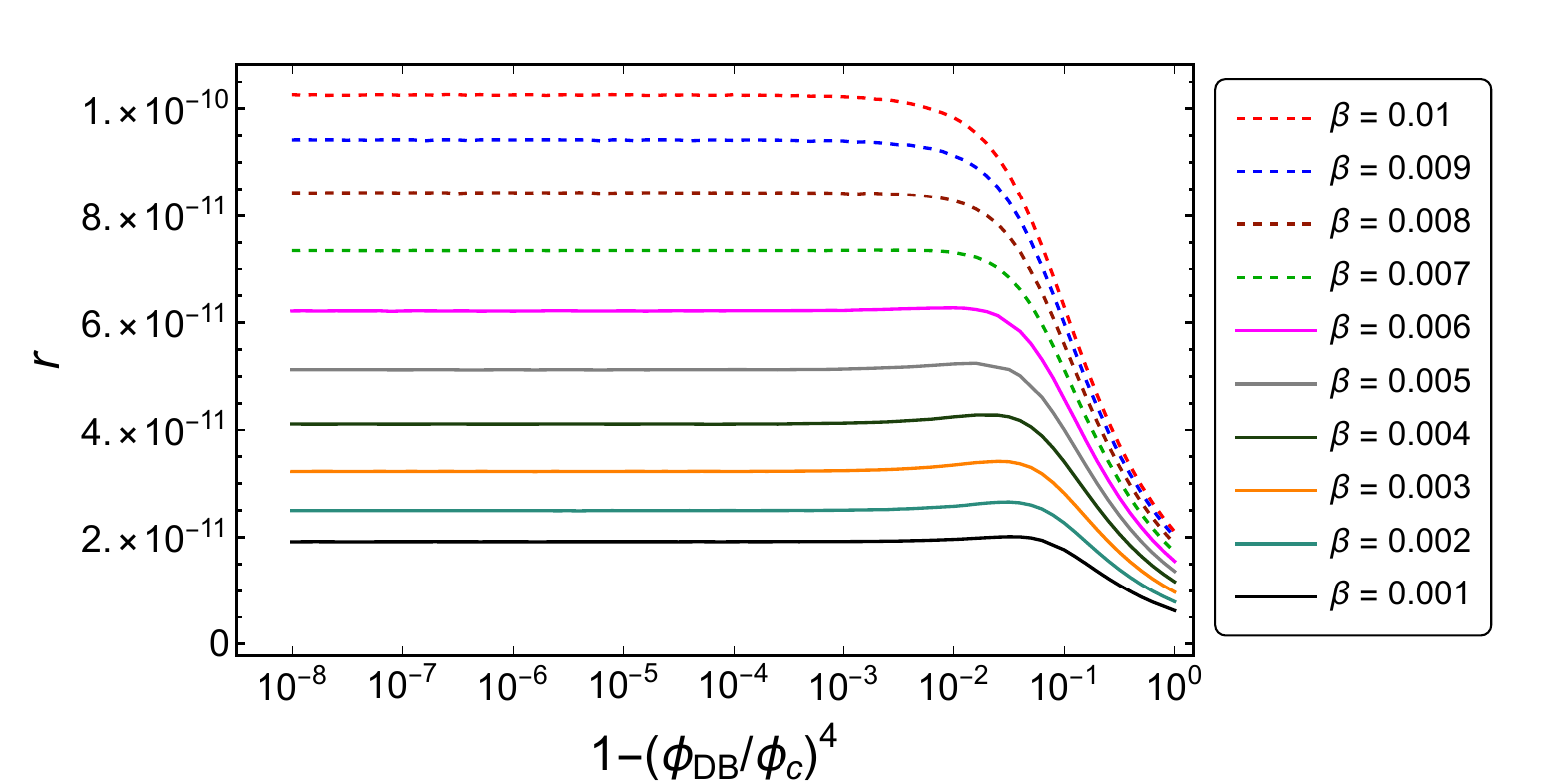}}
\caption{\label{fig:9}(a)\quad The scalar spectral index, $n_s$, for $(n=2,\,q=6)$ plotted against $(1-(\phi_{DB}/\phi_c)^4)$ for different values of $\beta$. (b)\quad The tensor-to-scalar ratio $r$, for $(n=2,\,q=6)$ plotted against $(1-(\phi_{DB}/\phi_c)^4)$ for different values of $\beta$. The parameter $\alpha$ in this case is negligibly small. Solid lines represent an increasing trend of $n_s$ with $\beta$ values, while dashed lines represent the decreasing trend.}
\end{figure*}

The scalar potential for $q=6$ is given by
\begin{equation}\label{I221}
\begin{split}
    V/V_{0}&\simeq \left(1-\left(\frac{\psi}{M}\right)^{6}\right)^{2}+2\left(\frac{\psi}{M}\right)^{6}\left(\frac{\phi}{\phi_{c}}\right)^{4}\\&
     +\alpha~\phi^{2} - \beta~\psi^{2}.
\end{split}
\end{equation}
The variation of $n_s$ with $\beta$ for the various values of $\alpha$ is depicted in Fig.~\ref{fig:8b}. For this analysis, the values of the fields have been fixed at the diffusion boundary. A decreasing trend in $n_s$ can be seen with increasing $\alpha$. For $\alpha \gtrsim 1/3$, the field $\phi$  becomes sufficiently heavy to quickly settle down to zero, and inflation proceeds primarily through the evolution of the $\psi$ field, a scenario that was explored in the context of the supersymmetric hybrid framework in  \cite{_eno_uz_2004}.

Our next objective is to investigate the impact of initial conditions. In Figure~\ref{fig:9a}, we present a plot of the scalar spectral index as it varies with field values along the diffusion boundary, considering different values of $\beta$, while keeping $\alpha$ approximately close to zero. 
This plot illustrates that within a certain range of field values, specifically when $(1-(\phi_{DB}/\phi_c)^6) < 10^{-2}$, the effect on the spectral index is negligible. Within a narrower range, $(10^{-2}\lesssim(1-(\phi_{DB}/\phi_c)^6)\lesssim10^{-1})$, there is a slight decrease in the value of $n_s$, which diminishes as field values become even larger, rendering the effect on $n_s$ insignificant once more. For very small values of $\phi$, the predictions converge to a single-field new inflation scenario discussed in \cite{_eno_uz_2004}. It's crucial to emphasize that, given the defined ranges for $\alpha$ and $\beta$, the current scenario highly favors the occurrence of successful new inflation that aligns with Planck data.
Moving on to Fig.~\ref{fig:9b}, we depict the tensor-to-scalar ratio $r$ against the boundary, using the same parameter values as in Fig.~\ref{fig:9a}. As is consistent with a standard small-field inflation model, the value of $r$ is exceedingly small. Solid lines represent an increasing trend of $n_s$ with $\beta$ values, while dashed lines represent the decreasing trend.

\subsection*{Reheating and Leptogenesis}

To discuss reheating and leptogenesis in the case with $n=2$ (or $q=6$), we replace $\Phi^2$ in $W$ with $NN$, with $N$ representing the right-handed neutrino superfield. In this general framework, we consider the inflaton to be the lightest sneutrino field $N_I \equiv N_1$. Reheating and leptogenesis can proceed here in a manner analogous to the models examined in \cite{Antusch_2005, Masoud_2021, Masoud:2021prr}.
Incorporating the relevant term in the superpotential (\ref{I1}) for reheating, along with the Yukawa term involving $N_i$, we arrive at the following expression,
\begin{equation}
W \supset   \kappa \lambda_{ij}  \frac{\Psi^m}{M_c^{m-1}}N_iN_j+Y_{ij}^{\nu}N_iL_jH_u.
\end{equation}
After the breaking of the underlying symmetry, the Majorana mass term for $N_i$ superfield, is determined as follows:
\begin{equation}\label{445}
 M^R_{ij}=\frac{\kappa \lambda_{ij}}{2}\left( \frac{M}{M_c}\right)^{m-1}M.
\end{equation}
From the above expression, the mass of the sneutrino inflaton $N_1$ is obtained as, 
\begin{equation}
M_R^I = \frac{\sqrt{2}\kappa M}{q+2} \left( \frac{M}{M_c} \right)^{\frac{q+2}{2} } <  M,
\end{equation}
assuming $\phi_c = M$. For $M_c = 10 M$ and $V_0 \sim 10^{-19} - 10^{-20}$, we obtain  $\kappa \simeq 0.01-0.032$ and $M_R^I \simeq 4.3 \times 10^{9}-1.4 \times 10^{10}$~GeV.

The inflaton undergoes decay through the Yukawa coupling, $Y_{1j}^{\nu}N_1L_jH_u$, resulting in the production of slepton and Higgs or lepton and Higgsino  with a decay width given by, 
\begin{equation}
     \Gamma_{N_I} \simeq \frac{y_{\nu}^2}{4\pi} M_R^I \simeq \frac{y_{\nu}^2}{4\pi} \frac{\sqrt{2}\,\kappa \, M}{q+2} \left( \frac{M}{M_c} \right)^{\frac{q+2}{2} },
\end{equation}
where $y_{\nu}^2=\left(Y_{\nu}Y_{\nu}^{\dag}\right)_{11}$. 
Following the end of inflation, the universe undergoes reheating as a result of inflaton's decay. The reheating temperature, $T_r$, given by 
\begin{equation}
T_r=\left(\frac{90}{\pi^2 g_*}\right)^{1/4}\sqrt{\Gamma_{N_I}},
\end{equation}
where $g_*=228.75$ for MSSM. Assuming a standard thermal history, we can establish a connection between the total number of e-folds, $N_0$, and the reheating temperature using the equation:
\begin{equation}
    N_0 \simeq 53+\frac{1}{3}\ln{\left[\frac{T_r}{10^9\text{GeV}}\right]}+\frac{2}{3}\ln{\left[\frac{V_0^{1/4}}{10^{15}\text{GeV}}\right]}.
\end{equation}

The lepton asymmetry generated by the inflaton decay undergoes a partial conversion into the observed baryon asymmetry through sphaleron processes, \cite{Kuzmin:1985mm,Fukugita:1986hr,Khlebnikov:1988sr}. In order to suppress the washout factor of lepton asymmetry, we assume $ M^{I}_R\gg T_r $. The baryon asymmetry can be expressed in terms of the reheating temperature as follows:
\begin{equation}
\frac{n_B}{n_{\gamma}} \lesssim 1.84 \left( \frac{3}{8\pi}\frac{\sqrt{\Delta m^2_{31}}}{v^2} \right) T_r,
\end{equation}
assuming a hierarchical structure of neutrino masses. Here, the atmospheric neutrino mass squared difference is $ \Delta m^2_{31} \simeq 2.6 \times 10^{-3}$~eV$^2$ and we take $\langle H_u \rangle \simeq v \simeq 174$~GeV  in the large tan$\beta$ limit. Importantly, it is worth noting that this expression is independent of the inflaton mas, $M_R^I$. Finally, the observed baryon-to-photon ratio $n_B/n_{\gamma}=(6.10\pm0.04)\times10^{-10}$ places a bound on reheating temperature, $T_r\gtrsim 10^6$~GeV. For our numerical calculations, we set $T_r =10^6$~GeV, a value sufficiently low to evade the gravitino overproduction problem.

\section{Conclusion}  \label{conclusion}

The idea of achieving a new inflationary phase within the context of waterfall hybrid inflation, as initially proposed in \cite{Clesse_2011}, has been extended to align with the observational data provided by the Planck mission. In this extension, we identify a single-field inflationary scenario that predominates in the large $\mu$ (phase-2), while a multifield inflationary scenario emerges in the regime characterized by relatively smaller $\mu$ values (phase-1). The initial conditions for this generalization are generated through valley hybrid preinflation. The predictions of inflationary observables are computed using the $\delta N$ formalism. 
Remarkably, the predictions for single-field scenarios with large $q$ values are observationally consistent, whereas multifield scenarios can also work well with smaller $q$ values, even as low as 6.

Implementation of this idea within a supersymmetric framework requires a generalized version of the tribrid inflation model. The inclusion of supergravity corrections, stemming from a non-minimal K\"{a}hler potential, further boosts the model's viability. As a specific example, we consider $n=2$ (or $q=6$) and replace the $\Phi$ field with the right-handed neutrino field $N$.  Notably, the interaction term between the waterfall-Higgs field and the matter field assumes a pivotal role in achieving successful reheating and enabling non-thermal leptogenesis.

\nocite{*}

\bibliography{NewInflation}
\end{document}